\documentclass[12pt,preprint]{aastex} 
\usepackage{enumerate}

\def\orchestra{\textsc{Orchestra}} 
\def\Msolar{\ifmmode M_{\odot}\else $ M_{\odot}$\fi}
\def\Mearth{\ifmmode M_{\oplus}\else $ M_{\oplus}$\fi}
\def\mearth{\ifmmode M_{\oplus}\else $ M_{\oplus}$\fi}

\def\mreduced{{\mu}}
\def\myC{{\tilde{C}}}
\def\myD{{\tilde{D}}}

\def\vesc{v_{\rm esc}}
\def\vforce{v_{\rm force}}
\def\vdestruct{v_{\rm dest}}
\def\vkep{v_{_{\rm K}}}

\def\tkep{T_{_{\rm\!K}}}
\def\tcirc{T_{_{\rm circ}}}

\def\tdamp{T_{\rm damp}}
\def\tgdamp{T_{g,\rm damp}}
\def\tprecbin{T_{\rm bin-pre}}

\def\Mp{M_p}
\def\Ms{M_s}
\def\Rp{R_p} 
\def\Rs{R_s}

\def\ain{{a_{\rm in}}}
\def\acrit{{a_{\rm crit}}}
\def\abin{{a_{\rm bin}}}
\def\ebin{e_{\rm bin}}
\def\apsprebin{{\dot{\varpi}_{\rm bin}}}
\def\tpreaps{{T_{\rm pre}}}
\def\eforce{e_{\rm force}}
\def\efree{e_{\rm free}}
\def\nbin{{\Omega_{\rm bin}}}

\def\rgc{{R_{\rm g}}}
\def\ngc{{\Omega_{\rm g}}}
\def\nsyn{\omega_{\rm syn}}
\def\kappae{\kappa_{\rm e}}
\def\kappai{\nu_{\rm i}}

\def\rp{r}

\def\SigmaOhXcgs{\left[\frac{\Sigma_0}{\textrm{\small 10\,g/cm$^2$}}\right]}
\def\SigmagOhMMcgs{\left[\frac{\Sigma_{g,0}}{\textrm{\small 2000\,g/cm$^2$}}\right]}
\def\aAU{\left[\frac{a}{\textrm{\small 1\,AU}}\right]}
\def\rkm{\left[\frac{r}{\textrm{\small 1\,km}}\right]}
\def\MMsun{\left[\frac{M}{\textrm{\small 1\,M$_\odot$}}\right]}
\def\dMhalfMsun{\left[\frac{\Mp-\Ms}{\textrm{\small 0.5\,M$_\odot$}}\right]}
\def\muhalfMsun{\left[\frac{\mreduced}{\textrm{\small 0.5\,M$_\odot$}}\right]}
\def\rhoIIIcgs{\left[\frac{\rho}{\textrm{\small 3\,g/cm$^3$}}\right]}

\def\abinOhpIIAU{\left[\frac{\abin}{\textrm{\small 0.2\,AU}}\right]}
\def\ebinOhpII{\left[\frac{\ebin}{\textrm{\small 0.2}}\right]}
\def\ainIIpVbin{\left[\frac{\ain}{\textrm{\small 2.5\,$\abin$}}\right]}

\def\ajIIAU{\left[\frac{a_{\rm j}}{\textrm{\small 2\,AU}}\right]}
\def\mjMJ{\left[\frac{M}{\textrm{\small 1\,M$_{\rm Jupiter}$}}\right]}

\def\kepler{\textit{Kepler}}
\def\lae{\lower 2pt \hbox{$\, \buildrel {\scriptstyle <}\over {\scriptstyle\sim}\,$}}
\def\gae{\lower 2pt \hbox{$\, \buildrel {\scriptstyle >}\over {\scriptstyle\sim}\,$}}

\begin{document}

\title{Planet formation around binary stars: 
Tatooine made easy}

\author{Benjamin C. Bromley}
\affil{Department of Physics \& Astronomy, University of Utah, 
\\ 115 S 1400 E, Rm 201, Salt Lake City, UT 84112}
\email{bromley@physics.utah.edu}

\author{Scott J. Kenyon}
\affil{Smithsonian Astrophysical Observatory,
\\ 60 Garden St., Cambridge, MA 02138}
\email{skenyon@cfa.harvard.edu}

\begin{abstract}

We examine characteristics of circumbinary orbits in the context of
current planet formation scenarios.  Analytical perturbation theory 
predicts the existence of nested circumbinary orbits that are
generalizations of circular paths around a single star. 
These orbits have forced eccentric motion aligned with the binary as well as
higher frequency oscillations, yet they do not cross, even in the
presence of massive disks and perturbations from large planets.  For
this reason, dissipative gas and planetesimals can settle onto these
``most circular'' orbits, facilitating the growth of protoplanets.
Outside a region close to the binary where orbits are generally
unstable, circumbinary planets form in much the same way as their
cousins around a single star.  Here, we review the theory and confirm
its predictions with a suite of representative simulations.  We then
consider the circumbinary planets discovered with NASA's
\kepler\ satellite. These Neptune- and Jupiter-size planets, or their
planetesimal precursors, may have migrated inward to reach their
observed orbits, since their current positions are outside of unstable
zones caused by overlapping resonances.  {\it In situ} formation
without migration seems less likely, only because the surface density
of the protoplanetary disks must be implausibly high.  Otherwise, the
circumbinary environment is friendly to planet formation, and we
expect that many Earth-like ``Tatooines'' will join the growing census
of circumbinary planets.

\end{abstract}


\section{Introduction}

Planet formation is robust. Transit detections by the \kepler\ satellite
\citep[e.g.,][]{borucki2011,how2012}, radial velocity campaigns
\citep{cum2008,how2010,zec2013,mayor2014}, direct imaging surveys
\citep{mac2014,tamura2014}, and gravitational lensing studies
\citep{gould2010} suggest that many if not all stars host planets
\citep{youdin2011b,dong2013}. To date, there are over 1500 confirmed
planets, and several thousand candidates \citep{exoplanets2014}.

The vast majority of the known planets orbit a single star 
\citep[e.g.,][]{mayor2011,cassan2012,burke2014,mullally2015}. 
Their formation in this setting is straightforward to describe, even 
if certain key details are not well understood 
\citep[e.g.,][]{saf1969,weth1980,gold2004,youdin2013}. The overall
process takes many small solid particles of dust and concentrates this
mass into a few large objects. Coagulation --- growth through sticking
or merging of planetesimals --- is driven by collisions.  Low relative
velocities favor growth, but can also slow it down. Higher velocities
can speed up the growth rate, but can also lead to destructive
collisions. Scattering and gravitational interactions pump up relative
velocities, while collisional damping and dynamical friction slow
things down.  A balance between these processes enables planets to
emerge from the dust.

In addition to the dozens of planets known to orbit one member of a
stellar binary \citep[e.g., $\alpha$~Cen~B; see][]{schneider2011},
a handful of planets are known to orbit both binary partners.  
Kepler-16b -- the first circumbinary planet discovered by the 
\kepler\ mission -- is a Saturn-mass planet at an orbital distance 
of about 0.7~AU \citep[][]{doyle2011}. 
Its binary host consists of a 0.7~\Msolar\ K-type star
and a 0.2~\Msolar\ red dwarf at an orbital separation of 0.22~AU.
Since then, six more ``Tatooines'' have been reported in the \kepler\ 
data set (see Table~\ref{tab:kepler}, below).  They are typically
Neptune- or Jupiter-size, and all orbit their hosts at distances
within roughly 1~AU. One binary, Kepler 47, hosts two such planets
\citep{orosz2012a}. A few more massive circumbinary planets are known
or suspected, but these objects orbit at much greater distances
compared with the binary semimajor axis
\citep[e.g.,][]{Beuermann2011}.

The circumbinary planets that orbit close to their hosts provide unique 
challenges for planet formation theory. The central stars strongly
perturb the region around them, clearing out orbits to distances of
2--5 times the binary separation \citep{holman1999}. Similarly,
circumstellar disks get eroded from the outside by the binary
partner. In either case,
secular excitations from the binary potential drive orbital          
crossings and destructive collisions
\citep{mori2004,meschiari2012,paarde2012,raf2013,lines2014}.  
Thus, planets may not be able to grow near their binary host.

To resolve this problem, planets may form at more remote distances,
where the time-varying part of the binary potential is weak.  To
arrive at their observed locations, they must then migrate through 
the circumbinary gaseous disk or scatter with neighboring gas giants
\citep[][]{pierens2008a,pierens2008b}.  While the simulations of this
process are compelling \citep[e.g.,][]{kley2014}, there are
uncertainties about starting conditions, typically a set of planetary
cores placed into a steady state disk. It is unclear if the
simultaneous growth of planetesimals and dissipation of the disk can
conspire to produce cores poised to migrate into their observed
orbital positions.

Toward understanding how circumbinary planets form, we re-examine a
fundamental issue: the nature of planetesimal orbits around binary
stars. Following the approach of \citet{lee2006} and
\citet{leung2013}, we describe a family of nested, stable circumbinary
orbits that have minimal radial excursions and never intersect.
While they are not exactly circular, these orbits play the same role
as circular paths around a single star. Gas and particles can damp to
these orbits as they dynamically cool, avoiding the destructive
secular excitations reported in previous work.  Thus planetesimals may
grow {\em in situ} to full-fledged planets.

We organize this paper to provide an introduction to the
Lee--Peale--Leung analytical theory of circumbinary orbits (\S2),
followed by numerical examples (\S3). We then discuss the role these
orbits play in planet formation (\S4), and close with a comparison to
observations (\S5), along with a summary and predictions of the ideas
presented here (\S6).

\section{The circumbinary environment}\label{sec:theory}

Planet formation relies on the ability of solid particles --- dust,
planetesimals, protoplanets --- to interact gently. Around a single
star, the family of nested (concentric) circular orbits offers this
possibility. Particles on coplanar circular orbits coexist without any
collisions. Gravitational interactions among particles (and with 
coexisting gas) induce random motions about these circular orbits,
which enables particles to merge into larger objects. Unchecked, 
gravitational interactions grow indefinitely and lead to destructive
collisions among particles \citep[e.g.,][]{gold1978}. Dynamical cooling
(through collisional damping, dynamical friction, or gas drag) is
essential \citep[e.g.,][]{gold2004,youdin2013}. When orbiting particles
cool, they need some common set of trajectories on which to settle. The 
family of circular orbits provides this non-intersecting, collisionless 
haven for particles in cold circumstellar disks.

A central binary dramatically alters these orbital dynamics
\citep[e.g.,][]{holman1999,musielak2005,pichardo2005,doolin2011}.
For a primary and secondary with comparable masses and binary 
eccentricity, $\ebin$, satellite orbits are unstable inside of 
a critical radius, $\acrit$, which is at least twice the binary 
separation, $\abin$. \citet{holman1999} derive an approximation 
for $\acrit$ from direct simulation of circumbinary particles:
\begin{eqnarray}\label{eq:acrit}
\acrit & \approx & 1.60 + 5.10\, \ebin - 2.22\, \ebin^2 + 
4.12\frac{\Ms}{\Mp+\Ms}
 -4.27\, \ebin \frac{\Ms}{\Mp+\Ms} 
\\ & &  - 5.09 \frac{\Ms^2}{(\Mp+\Ms)^2}  
+ 4.61\, \ebin^2 \frac{\Ms^2}{(\Mp+\Ms)^2},
\end{eqnarray}
where $\Mp$ and $\Ms$ are the masses of the primary and secondary,
respectively \citep[see also][]{pichardo2005,pichardo2008}. Inside 
this orbital distance, particles are cleared, creating a cavity 
around the binary.

Particles beyond the critical distance can be on stable, non-Keplerian 
orbits. In addition to their response to the central mass of the binary, 
these satellites also experience {\em forced} motion,
driven by the binary's time-varying potential.  This perturbation
prevents particles from maintaining circular or eccentric orbits. Instead,
particles may achieve a ``most circular orbit,'' defined as having the
smallest radial excursion about some guiding center, orbiting at some
constant radius $\rgc$ and angular speed $\ngc$ in the plane of the
binary \citep{lee2006,youdin2012}.  More generally, 
eccentric circumbinary orbits
may be composed of epicyclic motion about $\rgc$, as in the Keplerian
case, superimposed on the most circular orbit.  

In the rest of this section, we investigate existing analytic theory 
for circumbinary orbits and applications for planet formation. We also
include a brief discussion of instabilities and resonances, as a prelude
to numerical simulations in \S3.

\subsection{Analytical theory of circumbinary orbits}
\label{subsec:theory}

To describe satellite orbits about a central binary, we follow the
analytic theories of \citet{lee2006} and \citet{leung2013}, based on the 
restricted three-body problem \citep[see][]{szeb1967,murray1999}.  In 
this framework, a satellite's position and momentum come from equations 
of motion in the potential of a stellar binary.  This strategy differs
from previous work based on secular perturbation theory
\citep[e.g.,][]{hepp1978, marzari2000, mori2004,raf2013}.
In that approach, a satellite's osculating orbital elements, defined
with respect to a central point mass, evolve according to an
orbit-averaged disturbing function \citep[e.g.,][]{murray1999}.
However, these elements do not accurately track the orbits of the
\kepler\ circumbinary planets; the binary induces significant 
motion on dynamical time scales. Although modifications to the 
secular theory can accommodate this extra motion
\citep{georgakarakos2015}, any additional non-gravitational processes
like aerodynamic drag or planetesimal collisions are more easily 
described in terms of positions and momenta.  For these reasons
we adopt the approach of \citet{lee2006} and \citet{leung2013}.

The starting point of the Lee--Peale--Leung analysis is the
gravitational potential of the binary:
\begin{equation}
\label{eq:Phi}
\Phi = -\frac{G \Mp}{\sqrt{R^2+z^2+\Rp^2+2R \Rp \cos\Delta\phi}}
- \frac{G \Ms}{\sqrt{R^2+z^2+\Rs^2-2R \Rs \cos\Delta\phi}},
\end{equation} 
where $G$ is the gravitational constant and $\Delta\phi$ is angle
between the secondary and the satellite in a reference frame with the
binary's center of mass at the origin. In this frame, the massless satellite 
is at radial position $R$ in the plane of the binary and has altitude $z$
above this plane. The terms $\Rp$ and $\Rs$ denote the orbital distances 
of the primary and secondary.

To make headway, this potential is expanded in terms of the angle
cosines, converted from powers ($\cos^k(\Delta\phi)$) to multiple-angle
form ($\cos(k\Delta\phi)$).  For eccentric binaries, \citet{leung2013}
also expand the potential to first order in the binary eccentricity,
$\ebin$, using the epicyclic approximation to describe the variation
in binary separation and phase.  Then they seek solutions for the
excursion of the satellite from a guiding center on a circular orbit
of radius $\rgc$. The excursions in radial, azimuthal and altitude
coordinates are $\delta R$, $\delta \phi$, and $\delta z$ (which
is identically $z$, since all vertical motions are excursions from the
guiding center orbiting in plane of the binary). Their solution
can be estimated by writing the equations of motion and keeping only
terms linear in the perturbation coordinates and in the binary
eccentricity. Between the expansion of the potential and this
linearization, the problem reduces to the form of a simple, driven
harmonic oscillator.

To follow this prescription, we focus on motion in the binary's 
orbital plane. The potential is
\begin{eqnarray}
\label{eq:Phiexp}
\Phi & \approx & \sum_{k = 0}^{\infty}
\left\{ 
\Phi_{0k} \cos(k\Delta\phi) 
+ \ebin 
\left[ \Phi^e_{0k} \cos(k\Delta\phi)\cos(\nbin t)
\right.
\right.
\\ \nonumber
\  & \ & \ \ \ \ \ \ \ \ \ \ \ \ \ \ \ \ \ \ \ \ 
\left.
\left.
+\ \ 2 k \Phi_{0k} \sin(k\Delta\phi) \sin(\nbin t)
\right]
\right\}
\ \ \ \ \ \ \ \ \ \ \ [z = 0]
\end{eqnarray}
where $\nbin$ is the mean motion of the binary ($\nbin^2 =
G(\Mp+\Ms)/\abin^3$), and our choice of time $t$ fixes the orbital
phases of the binary and satellite. The potentials $\Phi_{0k}$ are
Fourier coefficients derived from the expansion of the potential in
terms of $\cos(\Delta\phi)$ evaluated at $\ebin=0$, while the
$\Phi^e_{0k}$ are those same coefficients giving the first-order terms
of a Taylor series in $\ebin$. 
Examples of these coefficients are
\begin{eqnarray}
\label{eq:PhiOO}
\Phi_{00} & = & -\frac{G M}{\rgc} -
\frac{G\mreduced}{\rgc}\left[
 \frac{1}{4}\frac{\abin^2}{\rgc^2}
   + \frac{9}{64}\frac{(\Mp^2+\Ms^2)}{M^2} \frac{\abin^4}{\rgc^4}
  + \frac{25}{256}\frac{(\Mp^4+\Ms^4)}{M^4} \frac{\abin^6}{\rgc^6}
 + ...\right]
\\
\Phi_{01} & = & -\frac{G\mreduced}{\rgc}\left[
  \frac{3}{8}\frac{(\Mp-\Ms)}{M}\frac{\abin^3}{\rgc^3} 
 + \frac{15}{64}\frac{(\Mp^4-\Ms^4)}{M^4}\frac{\abin^5}{\rgc^5}
 + ...\right]
\\
\Phi_{02} & = & -\frac{G\mreduced}{\rgc}\left[
  \frac{3}{4}\frac{\abin^2}{\rgc^2}
 + \frac{5}{16} \frac{(\Mp^3+\Ms^3)}{M^3}\frac{\abin^4}{\rgc^4}
 + \frac{105}{512} \frac{(\Mp^5+\Ms^5)}{M^5}\frac{\abin^6}{\rgc^6}
 + ...\right]
\\
\Phi_{03} & = & -\frac{G\mreduced}{\rgc}\left[
  \frac{5}{8} \frac{(\Mp-\Ms)}{M}\frac{\abin^3}{\rgc^3} 
 + \frac{35}{128} \frac{(\Mp^4-\Ms^4)}{M^4}\frac{\abin^5}{\rgc^5}
 + ...\right]
\\
\Phi_{04} & = & -\frac{G\mreduced}{\rgc}\left[
  \frac{35}{64} \frac{(\Mp^3+\Ms^3)}{M^3}\frac{\abin^4}{\rgc^4}
 + \frac{63}{256} \frac{(\Mp^5+\Ms^5)}{M^5}\frac{\abin^6}{\rgc^6}
 + ...\right]
\\
\Phi_{05} & = & -\frac{G\mreduced}{\rgc}\left[
 \frac{63}{128} \frac{(\Mp^4-\Ms^4)}{M^4}\frac{\abin^5}{\rgc^5}
 + ...\right];
\end{eqnarray}
and
\begin{eqnarray}
\Phi^e_{00} & = & -\frac{G\mreduced}{\rgc}\left[
  \frac{1}{2}\frac{\abin^2}{\rgc^2} +
 \frac{9}{16}\frac{(\Mp^3+\Ms^3)}{M^3}\frac{\abin^4}{\rgc^4} + 
 \frac{75}{128}\frac{(\Mp^5+\Ms^5)}{M^5}\frac{\abin^6}{\rgc^6} + 
...\right]
\\
\Phi^e_{01} & = & -\frac{G\mreduced}{\rgc}\left[
  \frac{9}{8}\frac{(\Mp-\Ms)}{M}\frac{\abin^3}{\rgc^3} + 
  \frac{75}{64}\frac{(\Mp^4-\Ms^4)}{M^4}\frac{\abin^5}{\rgc^5} + ...\right]
\\
\Phi^e_{02} & = & -\frac{G\mreduced}{\rgc}\left[
  \frac{3}{2}\frac{\abin^2}{\rgc^2} + 
  \frac{5}{4}\frac{(\Mp^3+\Ms^3)}{M^3}\frac{\abin^4}{\rgc^4} + 
  \frac{315}{256}\frac{(\Mp^5+\Ms^5)}{M^5}\frac{\abin^6}{\rgc^6} + ...\right]
\\
\Phi^e_{03} & = & -\frac{G\mreduced}{\rgc}\left[
  \frac{15}{8}\frac{(\Mp-\Ms)}{M}\frac{\abin^3}{\rgc^3} + 
  \frac{175}{128}\frac{(\Mp^4-\Ms^4)}{M^4}\frac{\abin^5}{\rgc^5} + ...\right]
\\
\Phi^e_{04} & = & -\frac{G\mreduced}{\rgc}\left[
  \frac{35}{16}\frac{(\Mp^3+\Ms^3)}{M^3}\frac{\abin^4}{\rgc^4} + 
  \frac{189}{128}\frac{(\Mp^5+\Ms^5)}{M^5}\frac{\abin^6}{\rgc^6} + ...\right]
\\
\Phi^e_{05} & = & -\frac{G\mreduced}{\rgc}\left[
  \frac{315}{256}\frac{(\Mp^4-\Ms^4)}{M^4}\frac{\abin^5}{\rgc^5} + ...\right].
\end{eqnarray}
where $M = \Mp+\Ms$ is the total mass and $\mreduced = \Mp
\Ms/(\Mp+\Ms)$ is the reduced mass\footnote{In some previous studies, 
  ``$\mu$'' is defined as the ratio of the secondary's mass to the total 
  mass.  Thus, our $\mreduced/M$ is equal to ``$\mu(1-\mu)$'' in 
  \citet{holman1999}, for example.}.  
The subscripts $jk$ designate that each term is measured in the plane of
the binary ($j=0$), and is the $k^{\rm th}$ harmonic as in
Equation~(\ref{eq:Phiexp}).  The missing terms are of order
$(\abin/\rgc)^7$, which can be as large as a percent in an idealized
system, and a fraction of a percent in observed binaries.

From the time-averaged potential ($\Phi_{00}$ in Equation~(\ref{eq:PhiOO})),
we obtain the angular speed of the guiding center, $\ngc$. It follows
from 
\begin{equation}
\label{eq:ngc}
\ngc^2  \equiv \frac{1}{\rgc} \left.\frac{d\Phi_{00}}{dR}\right|_\rgc
= 
  \frac{G M}{\rgc^3}\left\{1 + \frac{\mreduced}{M}\left[
    \frac{3}{4}\frac{\abin^2}{\rgc^2}  
     + \frac{45}{64}\frac{(\Mp^3+\Ms^3)}{M^3}\frac{\abin^4}{\rgc^4}
     + ... \right]\right\} ~ .
\end{equation}
The square root of $\ngc^2$ is the mean motion of the satellite.
The satellite's epicyclic and vertical frequencies are
\begin{eqnarray}\label{eq:kappae}
\kappae^2 & \equiv & \rgc \left.\frac{d\ngc^2}{dR}\right|_\rgc\!\! + 4\ngc^2
 = 
  \frac{G M}{\rgc^3}\left\{1 - \frac{\mreduced}{M}\left[
     \frac{3}{4}\frac{\abin^2}{\rgc^2}  
     + \frac{135}{64}\frac{(\Mp^3+\Ms^3)}{M^3}\frac{\abin^4}{\rgc^4}
     + ...\right]\right\}
\\
\kappai^2 & \equiv & \left. \frac{1}{z}\frac{d\Phi}{dz}\right|_{z=0,\rgc}
= \frac{G M}{\rgc^3}\left\{1 + \frac{\mreduced}{M}\left[
     \frac{9}{4}\frac{\abin^2}{\rgc^2}
     + \frac{225}{64}\frac{(\Mp^3+\Ms^3)}{M^3}\frac{\abin^4}{\rgc^4}
     + ...\right]\right\}
\end{eqnarray}
corresponding to the eccentricity and any motion out of the plane of
the binary \citep[for details regarding the motion out of the orbital 
plane, see][]{lee2006}.  In the limit that the binary separation
goes to zero, or when the binary mass ratio is extreme, both $\kappae$
and $\kappai$ become the Keplerian mean motion.

Forced oscillations experienced by a satellite depend on the synodic frequency,
\begin{equation}
\nsyn = \nbin - \ngc.
\end{equation}
In general, $\nsyn$ is just the average angular speed of the satellite
in a reference frame that rotates with the mean motion of the
binary. For a circular binary ($\ebin = 0$), the time varying force
felt by the satellite depends only on this frequency and its
harmonics.  In the case of an eccentric binary, the orbital frequency
of the binary also enters into the potential.

\subsubsection{Equations of motion}

Derivatives of the potential yield equations of motion in the excursion
coordinates ($\delta R$,$\delta\phi$,$z$). The strategy of
\citet{lee2006} is to cast these equations in the form of a forced
harmonic oscillator with natural frequencies $\kappae$ (for $\delta R$
and $\delta\phi$), $\kappai$ (for the $z$ coordinate) and driving
frequencies involving $\nbin$ and $\nsyn$.  The solutions in terms of
the full cylindrical coordinates are \citep[Equations (27), (31) and
  (35) in][]{leung2013}:
\begin{eqnarray}
\label{eq:mostcircr}
& & 
  R(t) = \rgc\left\{1 - \efree\cos(\kappae t+\psi_e) 
     - \sum_{k=1}^\infty {C_k}\cos(k\nsyn t)\right. 
\\ \nonumber
& & \ \ \ \nonumber
\left. - \ebin \left[\myC^e_0\cos(\nbin t) +
  \sum_{k=1}^\infty \myC^+_k \cos(k\nsyn t+\nbin t)
  +\myC^-_k \cos(k\nsyn t - \nbin t)
  \right]\right\}
\\ 
& & 
\label{eq:mostcircphi}
  \phi(t)  = \ngc\left\{ t + \frac{2\efree}{\kappae} 
    \sin(\kappae t+\psi_e) + 
     \sum_{k=1}^\infty \frac{D_k}{k\nsyn}\sin(k\nsyn t) \right.
\\
& & \nonumber
 \ \ \left. + \ebin \left[\frac{\myD^e_0}{\nbin}\cos(\nbin t) +  
    \sum_{k=1}^\infty \frac{\myD^+_k \sin(k\nsyn t+\nbin t)}{k\nsyn+\nbin}
+        \frac{\myD^-_k \sin(k\nsyn t-\nbin t)}{k\nsyn-\nbin}
    \right]\right\}
\end{eqnarray}
\begin{eqnarray}
\label{eq:mostcircz}
& &
z(t)  =   i \rgc \cos(\kappai t + \psi_i), 
\hspace{3.75in}
\end{eqnarray} 
where $\efree$ is the ``free'' eccentricity, $i$ is the inclination, and the
phase angles $\psi_e$ and $\psi_i$ are constants. The coefficients are
\begin{eqnarray}
  \label{eq:Ck}
  C_{k} & = & \frac{1}{\rgc(\kappae^2-k^2\nsyn^2)}
\left[\frac{d\Phi_{0k}}{dR}-\frac{2\ngc\Phi_{0k}}{\rgc\nsyn}
\right]_{\rgc}
\\
  \label{eq:Cpmk}
  \myC^e_0 & = & -\frac{1}{\rgc(\kappae^2-\nbin^2)}
  \left[\frac{d\Phi^e_{00}}{dR}\right]_{\rgc}
\\
  \myC^\pm_k & = & 
\frac{1}{\rgc\left[\kappae^2-(k\nsyn \pm \nbin)^2\right]}
\left[\pm k\frac{d\Phi_{0k}}{dR}-\frac{1}{2}\frac{d\Phi^e_{0k}}{dR}
- \frac{k\ngc(\pm 2k\Phi_{0k}-\Phi^e_{0k})}{R(k\nsyn \pm \nbin)}
\right]_{\rgc}
\\
  D_k & = & 2 C_k + \left[\frac{\Phi_{0k}}{\rgc^2\ngc\nsyn}\right]_{\rgc},
\\
  \myD^e_0 & = & 2 \myC^e_0  
\\
  \myD^\pm_{k} & = & 2 \myC^\pm_k + \left[ 
    \frac{k(\pm 2 k \Phi_{0k}-\Phi^e_{0k})}{2\rgc^2\ngc(k\nsyn\pm\nbin)}
  \right]_{\rgc} 
\end{eqnarray}
\citep[Equations (28--30) and (32--34) in][]{leung2013}.\footnote{The
coefficients $\myC$ and $\myD$ are equal to the corresponding $C$ and
$D$ in \citet{leung2013} divided by the binary eccentricity. For
example, $\myC^+_k = C^+_{k}/\ebin$ and $\myC^e_0 = C_0/\ebin$. Our
choice allows the dependence on binary eccentricity to appear
explicitly in the solutions of the excursion variables.}

\subsubsection{Components of the orbital motion}

From the \citet{leung2013} solutions, we see that circumbinary
orbits may be broken up into independent modes. The first mode 
is forced motion, prescribed by the characteristics of the binary 
and the orbital distance of the satellite from the center of mass. 
The second mode is ``free'' motion, fully analogous to eccentricity 
and inclination of orbits around a single central mass,
except that the epicyclic excursions are relative to an orbiting
reference frame locked into forced motion, as opposed to a circular 
guiding center.

The forced motion itself can be broken down into parts, including
fast, driven oscillations at the synodic frequency and the binary's orbital
frequency, plus the slower epicyclic oscillations operating at 
the orbital frequency, $\ngc$.  The higher frequency
contributions have radial excursion amplitudes that scale as
\begin{equation}
\left\{C_{2},\myC^\pm_{2}\right\} \times \rgc \sim  
\frac{\mreduced}{M}\left(\frac{\abin}{\rgc}\right)^5 \rgc
\end{equation}
to leading order in $\abin/\rgc$ \citep[e.g.,][]{leung2013}.
The slower epicyclic motion is associated with the forced eccentric orbit
\citep{hepp1978}. Its contribution to the radial excursions
is typically much larger, scaling as
\begin{equation}\label{eq:eforce}
\eforce \rgc = \myC^-_{1} \rgc \approx 
\frac{5}{4}\frac{\Mp-\Ms}{M} \ebin \abin.
\end{equation}

When the binary eccentricity is large, we rewrite the solutions to 
highlight the role of the forced eccentricity.  For example, the 
radial coordinate becomes
\begin{equation}
R(t) = \rgc \left[1 - \efree \cos(\kappae t+\psi_e) - \eforce\cos(\ngc t)
+ ....\right]
\end{equation}
This form of the solution is familiar from secular perturbation theory
\citep{murray1999}.

An important feature of the forced eccentric orbit is that it does not
precess. Its argument of periastron is locked in line with the
argument of periastron for the binary. All other components of the
forced motion are also synchronized to the binary's orbital frequency,
the synodic frequency or their harmonics. In this way, {\it the
  orbital paths of satellites experiencing only forced motion make a
  family of nested orbits that never intersect}.

In contrast to the forced eccentricity, the free eccentricity and
inclination are both associated with precession. This effect is a
direct result of the fact that the time-averaged potential around
a binary does not fall off as $1/R$ --- the binary's mass averaged over its
orbit is akin to an oblate spheroid \citep{hepp1978, murray1999,
  lee2006}.  The precession rates of the periastron and ascending node
are
\begin{eqnarray}\label{eq:efreeprecess}
\dot{\varpi} & = & \ngc-\kappae \approx \frac{3}{4}\frac{\abin^2}{\rgc^2}
   \frac{\mreduced}{M}\sqrt{G M/\rgc^3}
\\
\dot{\Omega}_{\rm node} & = & \ngc - \kappai
   \approx  -\frac{3}{4}\frac{\abin^2}{\rgc^2}
   \frac{\mreduced}{M}\sqrt{G M/\rgc^3}
\end{eqnarray}
and thus are approximately equal and opposite.

\subsection{Incorporating a gaseous disk}
\label{subsec:theorydisk}

The orbit solutions in
Equations~(\ref{eq:mostcircr})--(\ref{eq:mostcircz}) apply directly to
non-interacting particles in a disk with negligible mass. Now we
consider how gravity, pressure support, and aerodynamic drag from the
gas disk modifies these orbits.
To make headway, we assume that outside of the critical radius where
orbits are unstable, the binary's gravitational potential changes the
fluid flow in a disk from circular orbits to most circular paths,
similar to the satellite orbits described above, but dependent on the
physics of the disk.  Our premise is that fluid flow in a gas disk
generally tends to circularize, an effect that is enhanced by apsidal
precession of free eccentric orbits (see discussions in \citealt{lin1976},
\citealt{syer1992}, \citealt{ogilvie2001}, \citealt{ogilvie2014}, and
\citealt{barker2014}).  Thus nested, non-intersecting, most circular
streamlines are a natural extension to circular flows around a point
mass.

Continuing with this picture, we make the simplifying assumption that
to a good approximation the disk potential and gas pressure are
axisymmetric. If gas fluid elements travel on most circular orbits,
then radial and azimuthal variations in disk properties must exist.
However, they may be small; in the \kepler\ circumbinary systems, the 
forced eccentricities are $\eforce \sim 0.002$--0.044 (see
\S\ref{sec:kepler}). Analytical models confirm that close to the
binary, the gravitational effects stemming from disk eccentricity are
small compared with the stars' influence on circumbinary orbits.  In
this way, we build on previous descriptions of circumbinary disks that
assert strict axisymmetry in the disk by having fluid elements travel
in pressure-supported circular orbits
\citep[e.g.,][]{marzari2000,scholl2007}.

In the limit of a tenuous gas, this simple picture yields reasonable
results.  As pressure and disk gravity diminish, gas molecules damp 
to the same set of most circular orbits derived for satellites in 
\S\ref{subsec:theory}.  However, the model does not incorporate 
hydrodynamical effects, including viscosity, turbulence and 
gravitational instabilities, that may significantly modify disk 
structure \citep[e.g.,][]{pel2013}.
With this acknowledgement of the limitations of model,
we proceed to consider circumbinary orbits.

\subsubsection{Orbits in the presence of a massive disk}
\label{subsubsec:massdisk}

The gravity of a massive circumbinary disk modifies the orbits of
satellites around a binary star.  An axisymmetric disk, with a 
gravitational potential $\Phi_{d}(R)$ in the orbital plane of the 
binary, modifies the mean motion and the epicyclic frequencies of 
a circumbinary satellite. These properties of the satellite are
related to derivatives of the time-average potential; we derive
them by making the substitution
\begin{equation}
\Phi_{00} \rightarrow \Phi_{00} + \Phi_d
\end{equation}
in Equations (\ref{eq:ngc}) and (\ref{eq:kappae}).  These changes to
$\ngc$ and $\kappae$ cause only minor adjustments to most terms in the
orbit solutions except for the term associated with forced
eccentricity:
\begin{equation}\label{eq:eforcegas}
\myC^-_1 = \frac{1}{(\kappae^2-\ngc^2)}
\left[ - \frac{d}{dR}\left(\Phi_{01}+\frac{1}{2}\Phi^e_{01}\right)
+ \frac{1}{R}  \left(2\Phi_{01} - \Phi^e_{01}\right)\right]_{\rgc}.
\end{equation}
With $\ngc$ and $\kappae$ in the denominator, the magnitude of the 
forced eccentricity now has a wide range of values that depend on 
the form of the disk potential.  In plausible astrophysical conditions, 
\citet{raf2013} demonstrates that the denominator can be negative 
and large compared to a disk-free system.  The magnitude of the 
forced eccentricity then becomes small; the forced eccentric orbit 
is anti-aligned with the binary.

When the denominator in Equation~(\ref{eq:eforcegas}) goes to zero,
satellite orbits experience a secular resonance \citep{raf2013}.
Physically, this condition occurs when orbits with free eccentricity
do not precess; contributions to the apsidal precession from the disk
and from the binary are equal and opposite. To quantify this behavior, 
we choose a disk surface density of
\begin{equation}\label{eq:Sigma}
\Sigma_g(a) = \Sigma_{g,0} \frac{a_0}{a},
\end{equation}
with $\Sigma_{g,0} = 2000$~g/cm$^2$ and $a_0 = 1$~AU, typical of
observed gas disks \citep[e.g.,][]{dent2013}. The disk potential is
\begin{equation}\label{eq:diskpotential}
  \Phi_d = 2\pi G\Sigma_{g,0}\ a_0 \log(a/a_0)
\end{equation}
\citep[][Appendix A]{bk2011a}. The corresponding apsidal precession 
rate from the disk is
\begin{equation}
\dot{\varpi}_{d} \approx -\left[\frac{1}{2\ngc R^2}\frac{d}{dR}
  \left(R^2\frac{d\Phi_d}{dR}\right)\right]_{\rgc} = 
-\frac{\pi}{\ngc\rgc} G \Sigma(\rgc).
\end{equation} 
Setting the magnitude of this expression equal to the apsidal
precession rate from the binary (Equation~(\ref{eq:efreeprecess}))
yields the radial position of the resonance. 

For the \kepler\ circumbinary planets, the resonance lies at an
orbital distance of roughly 2~AU.  
%
%
This distance is between the orbits
of the known planets \citep{raf2013} and the likely position of the
snow line \citep{kk2008}.  As the disk dissipates, however, the
resonance sweeps outwards past the snow line. This motion may have
interesting implications for the formation and inward migration of gas
giants.\footnote{It is beyond our scope to treat the orbital dynamics
  at the resonance in detail. Aside from its impact on the orbits of
  solids, it is unclear how the gas reacts.  Within the
  Lee--Peale--Leung theory, the evolution equation for the $\myC^-_1$
  mode on resonance is an undamped, driven harmonic oscillator with a
  natural frequency of $\ngc$.  This solution predicts a linear growth
  time scale for eccentricity of $10^3$--$10^4$ years for the
  \kepler\ circumbinary planetary systems, which may be an
  overestimate \citep[][Figure 6]{meschiari2014}.}

An extension to the axisymmetric form for the disk potential would 
account for the non-axisymmetric ebb and flow of fluid elements on most
circular orbits. A starting point is to approximate the streamlines
with the forced eccentric orbits from secular perturbation theory
\citep[e.g.,][]{silsbee2015a, silsbee2015b}.  The gravitational field
of the eccentric disk then affects the forced eccentricity, but not 
the apsidal alignment \citep[see Equations (19)--(21) in][]{silsbee2015b}. 
We could then calculate a self-consistent value for $\eforce$ where the 
orbits of the fluid elements follow paths that they help to generate.  
%
%
Further extensions to the 
theory of eccentric orbits would include the effects of geometric 
compression of fluid density from the continuity equation and related 
hydrodynamical effects for flow along most circular paths 
\citep[e.g.,][]{ogilvie2014, barker2014}.

\subsubsection{Orbits in the presence of a pressurized disk}
\label{subsec:pressurizedgas}

The pressure in protoplanetary gas disks modifies the orbit of any
fluid element. In a simple axisymmetric circumstellar disk where 
the radial pressure gradient is positive, pressure support leads to 
sub-Keplerian orbits \citep[e.g.,][]{weiden1977a,birn2010,chiang2010}.  
The mean orbital speed of the gas relative to a circular Keplerian 
orbit is
\begin{equation}\label{eq:eta}
\left|\Delta v\right| \approx \eta \rgc \ngc,
\end{equation}
where $\eta \sim 10^{-3}$ \citep[e.g.,][]{weiden1977a}. In a circumbinary
disk, we estimate the effect of pressure by assuming that the
pressure gradient is radial, and treating $\eta$ as a constant,
independent of orbital position, at least in some local region of the
disk.
We then may let pressure support appear
in the equations of motion as a change in the total mass of the binary:
\begin{equation}
\Phi_{00} \rightarrow (1-2\eta) \Phi_{00},
\end{equation}
leading to small modifications in quantities including $\ngc$ and
$\kappae$. Thus, gas pressure increases the forced eccentricity by
a factor of $1+2\eta$ compared to an unpressurized disk.  

In this approximation, the gas and solids not coupled to it are on
distinct most circular orbits \citep{thebault2006, meschiari2014,
silsbee2015b}. Nonetheless, the forced eccentric orbits of disks
with and without pressure support are apsidally aligned.  Furthermore
the differences between forced eccentric paths of gas and the
uncoupled solids are small. They lead to relative {\em radial} speeds
between gas and these solids of
\begin{equation}
\Delta v_r \sim \eta \eforce \vkep,
\end{equation}
which, for anticipated values of $\eforce$ for the
\kepler\ circumbinary planets, is roughly two orders of magnitude
smaller than the ``headwind'' felt by the solids as they plow through
the gas.  In other words, the most circular paths associated with
pressure-supported gas and uncoupled solids are sufficiently similar
that relative velocities arise predominantly from their azimuthal
motion, just as in circumstellar disks.

\subsubsection{Gas drag}
\label{subsec:gasdrag}

When the gaseous disk has internal pressure, material orbits the 
central object more slowly than the solids. The solids then
experience aerodynamic drag \citep{ada76,weiden1977a}. Particles 
smaller than about a centimeter are fully entrained in the gas.
Planetesimals with radii of 1~km or more barely feel the gas.
Both sets of particles follow distinct most circular orbits.  
For intermediate particle sizes, aerodynamic drag slows azimuthal 
speeds. Without the benefit of radial pressure support, these 
objects spiral inward (if the gas is sub-Keplerian).  The inspiral 
time scales are as fast as $1/\eta$ dynamical times 
\citep{ada76,weiden1977a,weiden1993,chiang2010,youdin2013}.

To describe the dynamics of a particle experiencing gas drag, we take
advantage of the small difference between the forced eccentricities of
a pressurized gas disk and satellites that orbit in the absence of gas
drag. Approximating the effect of gas as a constant azimuthal
headwind, the equations of motion resolve to a most circular orbit
with some intermediate forced eccentricity and a small amount of
radial drift \citep[e.g.,][]{youdin2013}. Importantly, if the
epicyclic motion of the entrained particles and large solids are
apsidally aligned, then so are the epicyclic orbits of these
intermediate-size bodies. We confirm below using numerical tests 
that this description is reasonable 
(\S\ref{sec:numerics}).

\subsection{Significance for circumbinary planet formation}

Here, we highlight several features of the orbit solutions described
in \S\ref{subsec:theory} and modifications arising from the presence 
of a gas disk (\S\ref{subsec:theorydisk}):

\begin{itemize}
\item The forced epicyclic motion (the $\myC^-_{1}$ term) is
  synchronized with the guiding center. The addition of an
  axisymmetric potential, whether it describes the effects of a
  massive disk or mimics the behavior of radial pressure, preserves
  this relationship. In general, forced eccentric orbits remain 
  apsidally aligned with the binary.  In a disk free of gas 
  pressure, particles on these paths never collide.

\item Pressure support changes the most circular paths of the gas
  streamlines relative to the orbits of solid particles not susceptible 
  to aerodynamic drag. This behavior generates relative epicyclic 
  motion between the gas and the solids. If the forced eccentric 
  orbits are apsidally aligned with the binary, however, then the 
  relative speeds induced by the differences in the most circular 
  paths are much smaller that the usual headwind felt by the solids 
  plowing through the gas. 

\item The forced epicyclic motion is coupled in its alignment with the
  binary. Thus if the binary itself precesses slowly, then the
  satellite's argument of periapse also precesses. To visualize this 
  point, we note that a slowly precessing binary orbit has slightly 
  different radial and azimuthal frequencies. To the solutions of
  \citet{leung2013}, we thus add the precession rate $\apsprebin$ to
  the binary's orbital frequency, $\nbin$, without modifying the
  synodic frequency $\nsyn$ between satellite and binary. The terms
  most greatly affected by this small change are the ones associated
  with the forced eccentricity (e.g., with $\myC^-_1$). The time
  dependence in these terms then transforms as
\begin{equation}
\cos(\nsyn t-\nbin t) \rightarrow \cos((\ngc+\apsprebin)t).
\end{equation} 
Thus the satellite's forced epicyclic motion precesses with the
binary, remaining apsidally aligned (in secular resonance).

\end{itemize}

These properties may be essential for circumbinary planet formation.
They suggest that most circular orbits remain nested and
non-intersecting even in the presence of a massive disk or if the
binary precesses due to interactions with a massive disk or a distant
planetary/stellar perturber.  Gas dynamics add complications, although
if a protoplanetary gas disk is nearly axisymmetric and streamlines
follow most circular orbits, then the presence of the binary induces
only small additional relative speeds compared to the circumbinary
case.  Thus issues of entrainment, gas drag, and the ``one-meter
barrier'' from circumstellar planet formation 
\citep[e.g.,][]{youdin2004a,birn2010,chiang2010,windmark2012,garaud2013,youdin2013}
carry over to the circumbinary environment.

Other studies of circumbinary planet formation include disks that are
exactly axisymmetric and circular in their orbital flow
\citep[e.g.,][]{scholl2007} or that are neither axisymmetric or
apsidally aligned with the binary \citep{silsbee2015b}. In these
cases, aerodynamic drag on particles leads them to achieve orbits that
depend on their physical size. In misaligned disks, for example, the
magnitude of the forced eccentricity and the apsidal orientation of
forced eccentric orbits depend on particle size \citep{silsbee2015b}.
In either case, particles of different size experience high-speed,
destructive collisions \citep{marzari2000,thebault2006}. However, gas
disks are strongly dissipative. We assume that most circular orbits,
aligned with the binary, are a good first approximation to the paths
of fluid elements.  In any event, as the gas dissipates, we expect
that the fluid elements and solids are likely to settle on the same
set of most circular orbits.

\subsection{Summary of the analytical theory}

The analytic theory of circumbinary orbits predicts a family of nested
most circular paths. Satellites orbiting with these paths (i) make
well-defined minimal radial excursions and (ii) never collide.  Thus,
these paths define ``dynamically cold'' orbits in exactly the same way
as circular orbits around single stars.  As in the Keplerian case,
satellites may have additional ``free'' eccentricity and inclination
to describe motion about these paths. For any
combination of free and forced eccentricity, the dynamics of
planetesimals as they stir or damp each other takes place in the frame
of these most circular paths.

While the physics of gaseous protoplanetary disks is uncertain, gas fluid 
elements can follow most circular paths, even when the disk mass and
pressure support are significant.  These orbits serve as reference
frames for local hydro- and aerodynamics. We expect that gas settles
to these orbits, since they allow for streamlines to be nested and
non-crossing with minimal radial excursions, despite any forced
eccentric motion.  Studies of eccentric gas disks around a single star
\citep{syer1992, ogilvie2001, ogilvie2014, barker2014} highlight
potential differences between the hydrodynamics of circular flows and
eccentric ones, including mass conservation along eccentric
streamlines \citep{ogilvie2014, barker2014}. In circumstellar disks,
gas might find stable orbits with some free eccentricity
\citep{syer1992}.  However, it seems more likely that apsidal
precession circularizes these orbits \citep{ogilvie2001}. For
circumbinary disks, we expect that apsidal precession leads to fluid
flow along most circular paths.

\subsection{Limitations of the theory: resonances and chaos}
 
Limitations of the theory stem from its perturbative approach,
particularly that it is linear in eccentricities.  Aside from
singularities in the coefficients in the solutions (e.g., $C_{k}$,
$\myC^\pm_k$) at the 1:1, 1:2 and 1:3 commensurabilities --- along
with the possible resonance induced by a massive disk \citep[see
  \S\ref{subsubsec:massdisk}, above]{raf2013} --- the
theory does not accommodate resonant effects \citep{wisdom:1982,
  wisdom:1983, lecar2001}.
These phenomena would arise in the equations of motion if the
satellite's eccentric motion were explicitly included in the
calculation of the force. Then, solutions may become chaotic and
possibly unstable \citep{wisdom1980}.

The formula for $\acrit$ provides a starting point for investigating
instability \citep[Equation \ref{eq:acrit};][]{holman1999}. In this
approximation, the location of unstable orbits close to the binary
varies smoothly with the mass ratio and binary eccentricity. However,
locating all of the unstable orbits is more complicated
\citep{musielak2005,doolin2011,chavez2015}. 
Overlapping resonance conditions generate instability
\citep{chirikov1959,chirikov1979,wisdom1980}; thus, unstable orbits
can exist in narrow, isolated ranges of orbital distance $a$ beyond
$\acrit$. In the binary Kepler-16 (Table~\ref{tab:kepler}), the 5:1
resonance, located outside of $\acrit$, is unstable. Although it is
weaker, the 6:1 resonance is also unstable
\citep[e.g.,][]{pierens2007}.  In between these resonances, orbits are
stable; Kepler-16b resides on one of these stable orbits
\citep{popova2013,chavez2015}. Beyond the 6:1 resonances, all
orbits are stable for small planetary eccentricities. When we discuss formation
mechanisms for circumbinary planets, we return to this issue
(\S\ref{sec:kepler}).

\section{Numerical simulations}\label{sec:numerics}

The analytic theory for circumbinary orbits is first-order accurate in 
the binary 
eccentricity and is derived from equations of motion linearized in the 
excursions away from the guiding center. Despite these limitations, it 
compares well in numerical experiments even when the binary eccentricity 
is moderately large. We summarize these experiments in this section.

\subsection{Code description}

We apply our planet formation code \orchestra\ \citep{bk2006,bk2011a}
to compare with the theoretical results in
\S\ref{sec:theory}. \orchestra\ is a hybrid $n$-body--coagulation code
for tracking the emergence of individual planets ($n$-bodies) from a
sea of smaller particles that can be characterized statistically (the
coagulation ``grid'' with bins for radial position and particle mass).
In a standard hybrid calculation, small particles within a set of
concentric annuli begin with an adopted radial surface density and
initial $e$ and $i$ relative to a circular orbit \citep[see
  also][]{saf1969,liss1987,spaute1991,weth1993,weiden1989,kl1998,kb2008}.
Initially, there are no $n$-bodies. As massive objects evolve in the
grid, the most massive are ``promoted'' into the $n$-body part of the
code. The subsequent evolution of the $n$-bodies and the grid are
linked together, enabling the simultaneous tracking of gravitational
interactions and collisions that lead to accretion, merging and
fragmentation \citep[e.g.,][]{kb2006,kb2010,kb2014}.

\orchestra\ has other capabilities which allow us to consider a
variety of problems in planet formation and evolution.  It can track a
swarm of massive or massless tracer particles -- usually sampled from
the orbital distribution of the coagulation particles -- to mediate
interactions that involve resonances and migration \citep{bk2011b}.
The code also includes interactions with a massive gas disk, both
through gravity \citep{bk2011a} and aerodynamic drag
\citep{weiden1977a,kb2002a}.  Tests of these and other elements of the
code, including our choice of time integrator --- either symplectic
\citep{yoshida1990} or adaptive Richardson extrapolation at 6$^{\rm
  th}$-order \citep{bk2006} --- are summarized in \citet{kb2001},
\citet{bk2006}, \citet{bk2011a}, and \citet{kb2015}.

Here we use the $n$-body component of the code with tracer particles
to calculate orbits around a binary. The primary and secondary stars
are $n$-bodies, evolved with the 6$^{\rm th}$-order symplectic
integrator.  Energy errors are better than one part in $10^{10}$.  In
some runs (see below) we include a third, Jupiter-mass $n$-body as a
perturber. In others we include a massive gas disk, choosing a surface
density $\Sigma$ in Equation~(\ref{eq:Sigma}) and the corresponding
gravitational potential in Equation~(\ref{eq:diskpotential}) to use in
the equations of motion for the tracers. This approach allows us to
test the analytic theory in \S\ref{sec:theory} using orbit solutions
derived numerically at high accuracy.

\subsection{Simulation results}

Simple $n$-body experiments with \orchestra\ allow us to test
several key features and predictions of the analytic theory (\S2).
For these studies, we use the Kepler-16 system as an example
\citep{doyle2011}, adopting binary parameters in Table~\ref{tab:kepler} 
as estimated by \citet{leung2013}.  

We begin with several illustrations of circumbinary orbits with no
other massive perturbers. Our first example focuses on most circular
orbits, showing both high-frequency oscillations and epicyclic
motion. In a second example, we consider the difference between the
time evolution of particles on most circular and `initially' circular
orbits around a eccentric binary.  Particles on most circular orbits
do not precess. Particles which start out on geometrically circular
orbits have some free eccentricity; this component of the orbit
precesses. By selecting an initially circular orbit with equal parts
of free and forced eccentricity, we show how the precession of the
free eccentric orbit modifies trajectories around the binary.

This second example identifies issues with previous $n$-body studies
of circumbinary planet formation \citep[e.g.,][]{mori2004, scholl2007,
  meschiari2012, paarde2012}.  In these analyses, initially
geometrically circular orbits precess, causing high relative collision
velocities that can inhibit the growth of planetesimals into
planets. Most-circular orbits have smaller radial excursions and
provide a calmer frame of reference for planet formation.

We then explore the impact of a massive disk, gas drag, and a
Jupiter-mass perturber. Our goal is to investigate whether most circular
orbits remain ``most circular'' in the context of a protoplanetary
disk around a binary star.

\begin{figure}[htb]
\centerline{\includegraphics[width=7.0in]{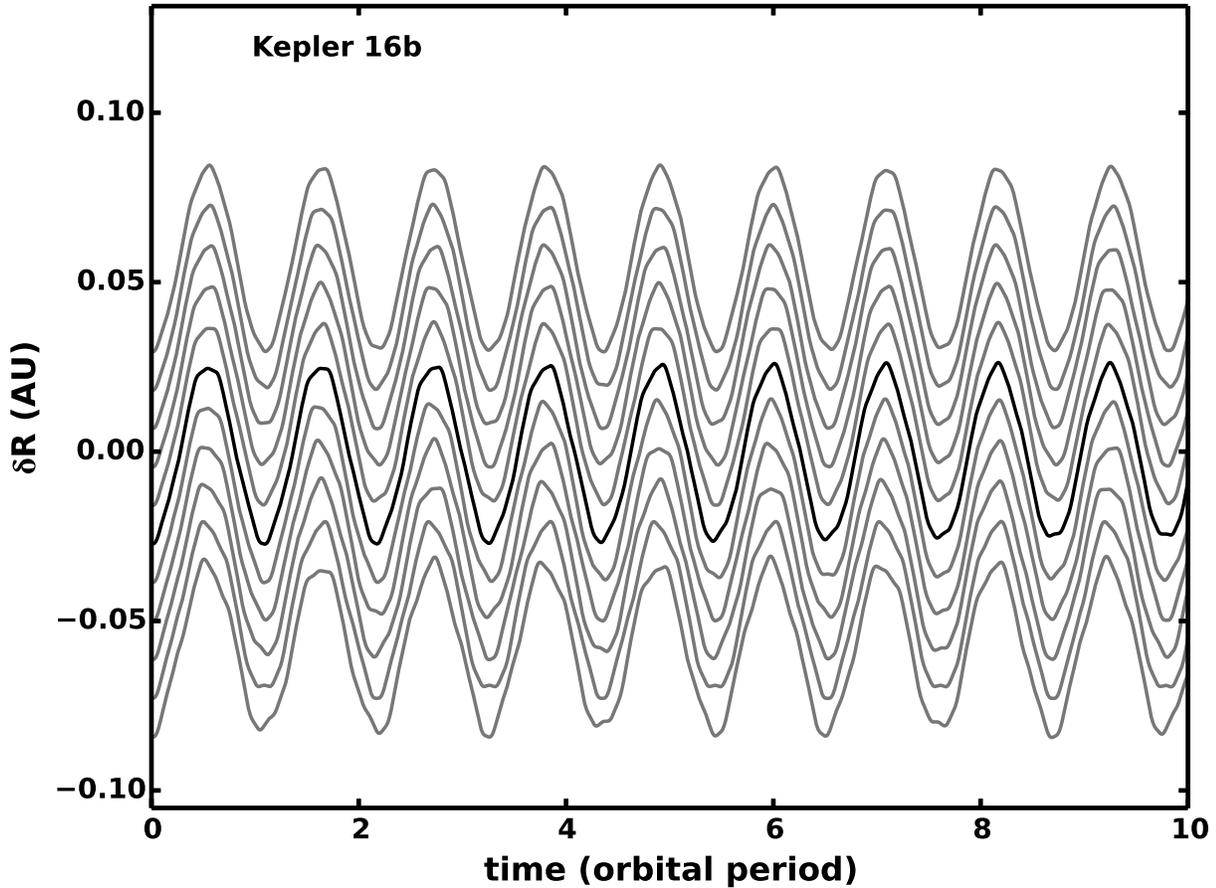}}
\caption{\label{fig:kepecckep16} 
The radial excursion of Kepler-16b on most circular orbits.
The dark curve shows the radial excursion of a satellite at the
orbital position of the planet in the absence of free eccentricity
and inclination, plotted as a function of orbital phase (in units
of the planet's orbital period). The gray curves show orbits at
slightly displaced orbital distances. The trajectories are dominated
by the forced eccentric orbit.  Comparatively small 
higher-frequency oscillations are visible in the curves. Despite
the appearance that these oscillations are not in phase between
the curves, the orbits do not cross.}
\end{figure}

\begin{figure}[htb]
\centerline{\includegraphics[width=7.0in]{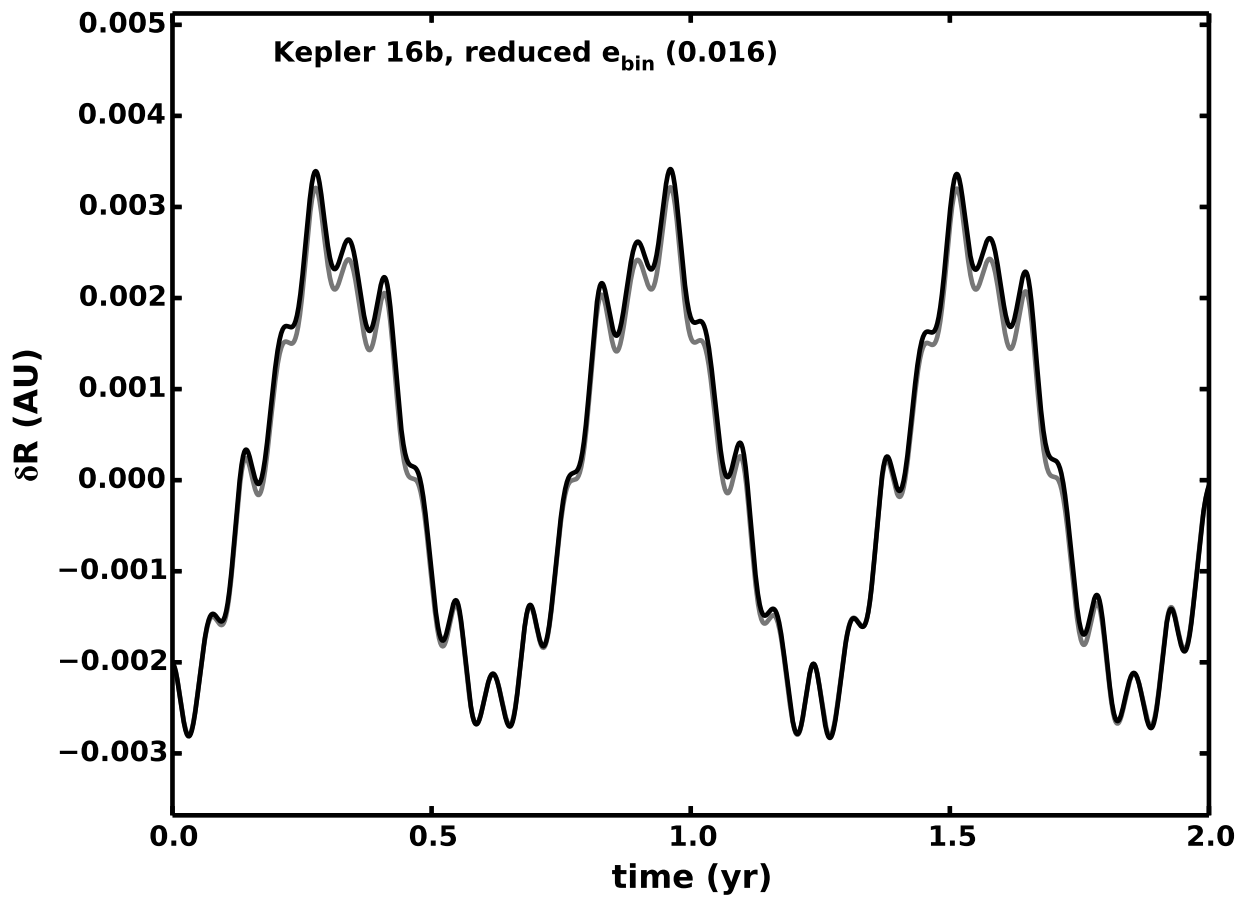}}
\caption{\label{fig:kepecckep16elow} Illustration of radial excursions
  of a satellite on a most circular orbit. The orbital configuration
  and binary masses are derived from Kepler-16 and its planet
  Kepler-16b at $a = 0.70$~AU.  The binary eccentricity has been
  reduced by a factor of 10 from the real system; the plot then
  distinguishes the forced eccentricity (larger amplitude, driven 
  at the orbital period, $\sim 0.6$~yr) and the high-frequency
  oscillations (smaller amplitude driven at the binary's orbital
  period and the synodic period of the satellite relative to the
  binary, $\sim 0.1$~yr). The black curve is from simulation while
the gray curve is from analytical theory (Equation~(\ref{eq:mostcircr})).
}
\end{figure}

\begin{figure}[htb]
\centerline{\includegraphics[width=7.0in]{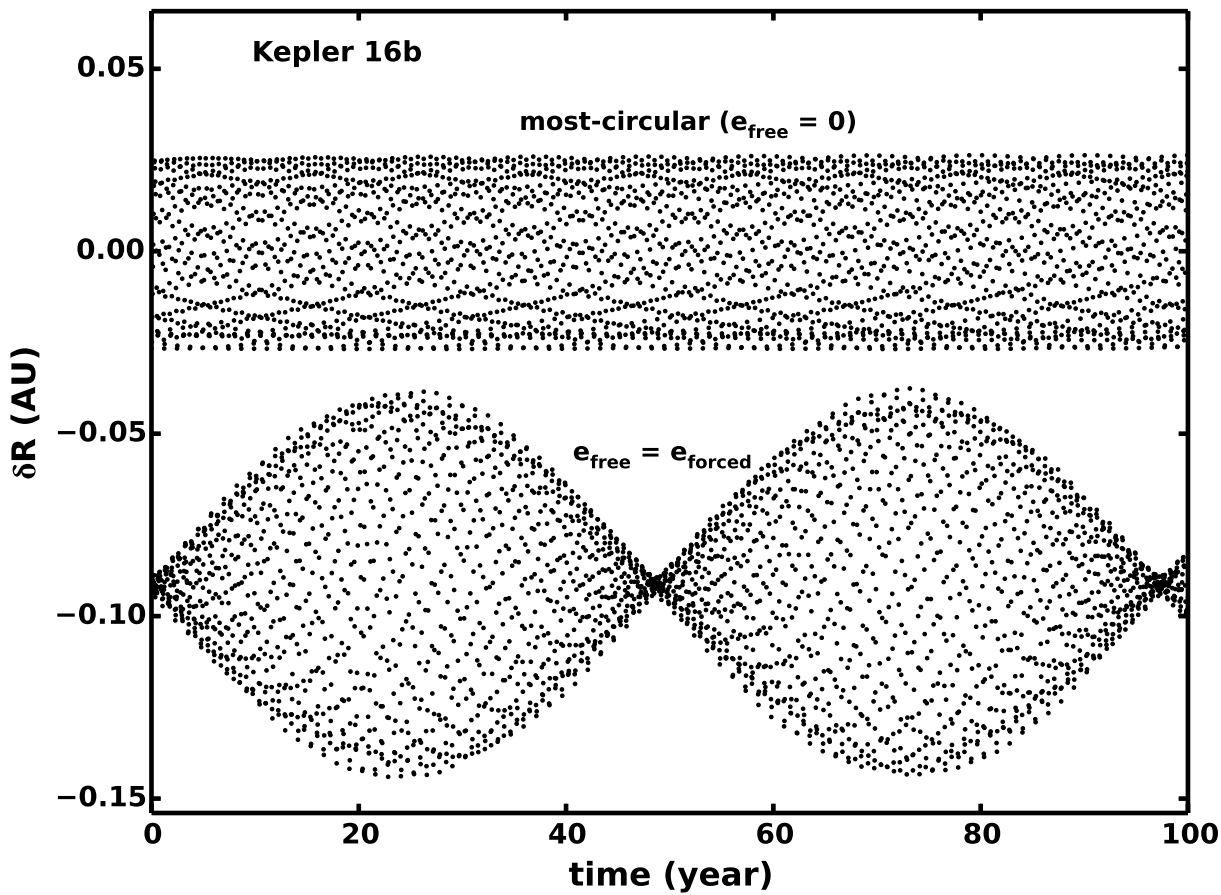}}
\caption{\label{fig:kepecckep16x} Comparison between a most circular
  orbit and one initially on the circular path of the guiding center
  (zero eccentricity). The most circular path (upper curve) is
  non-precessing, with excursions from the guiding center that have a
  maximum amplitude that does not drift over time.  In contrast, a
  particle that is set up on a ``circular'' orbit --- launching the
  particle from its guiding center with the speed of uniform
  circular motion about the binary center of mass --- has a mixture of
  free and force eccentricities in equal measure. The relative phase 
  allows zero eccentricity at the start. The result (lower curve, 
  displaced from the upper one for clarity) is the beat pattern with 
  a frequency given by the precession rate of
the free eccentric orbit (Equation~(\ref{eq:efreeprecess})).}
\end{figure}

\begin{figure}[htb]
\centerline{\includegraphics[width=7.0in]{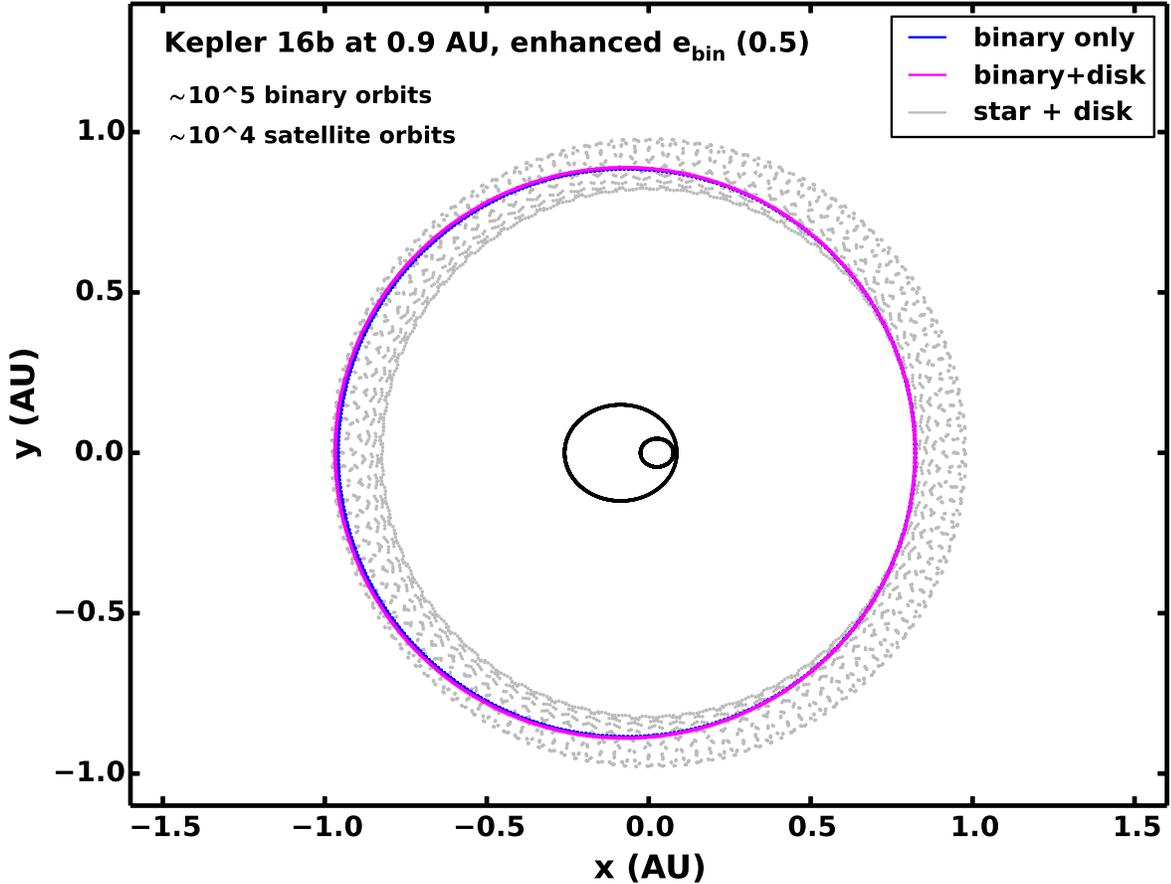}}
\caption{\label{fig:kepecckep16ehixgrav} Simulated most circular
  orbits around a binary with moderately high eccentricity. The
  orbital configuration is derived from Kepler-16, as in the previous
  figures, but the binary eccentricity is set to $0.5$, and the
  satellite is placed at 0.9~AU, just beyond the critical radius for
  stability ($\acrit=0.86$~AU). The center of mass of the system is at
  the origin in this $x-y$ map of the plane of the binary. The
  secondary's pericenter is on the positive $x$-axis. The satellite
  (blue dots) tracks a narrow elliptical path with an eccentricity of
  $\eforce \approx 0.08$, fixed and aligned with the binary for $\sim
  10^4$ orbital periods ($\sim 10^5$ binary orbits). When the
  potential of a massive disk ($\Sigma$ of 2000~g/cm$^2$ at 1~AU) is
  included, the satellite's most circular path does not precess 
  (magenta points, mixed in with the blue ones). For reference, 
  we show samples of a satellite orbiting a single star with Kepler-16's 
  total mass, also in this disk potential (gray points). In this case, 
  the precession of the satellite's argument of periastron is rapid 
  ($\sim 0.005$~rad/yr); the points fill an annular swath. In these 
  simulations, there is no interaction between the disk and the binary.}
\end{figure}

\begin{itemize}
\item{\em A sequence of most circular orbits.}
  Figure~\ref{fig:kepecckep16} shows the result of test particles on
  most circular paths near the orbital distance of Kepler-16b.  In
  these orbits, most of the motion comes from the forced eccentricity.
  As suggested by the Figure, the orbits are all nested and do not
  intersect.

\item{\em A most circular orbit in detail.}
  Figure~\ref{fig:kepecckep16elow} contains radial excursions of a
  satellite of Kepler-16 where the binary's eccentricity
  is reduced by a factor of ten. With this low eccentricity, both the
  high-frequency driven oscillations and the forced eccentric orbit are
  apparent.  This Figure shows a direct comparison to theory
  (Equation~(\ref{eq:mostcircr})).

\item{\em ``Circular'' versus Most-circular.}  Satellites can be
  launched on paths that are initially more circular (smaller radial
  excursions) than the most circular orbits discussed here. However,
  these satellites do not remain on circular paths as their orbits
  evolve over time.  Figure~\ref{fig:kepecckep16x} illustrates this
  point. For a central binary with non-zero $e$, an orbit with equal
  parts of free and forced eccentricity can be initialized on a purely
  circular orbit about the binary center-of-mass. At the start (time =
  0 in the figure), this orbit has very little epicyclic motion (e.g.,
  $\delta R$ is close to zero).  Over time, the free eccentric orbit
  precesses (Equation~(\ref{eq:efreeprecess})); the forced eccentric
  orbit does not.  Hence the two modes drift in phase, causing the
  beat pattern in the radial excursion shown in the Figure. Thus,
  orbits can be geometrically ``circular'' but only
  temporarily. Most-circular orbits have the smallest radial
  excursions over the long term.

  Previous studies, including our own \citep{kb2014}, describe
  simulations to track particle dynamics around binaries. If particles
  are set up initially on geometrically circular orbits (where free
  and forced eccentricities cancel at $t=0$), they appear to
  experience ``secular excitations'' where the total eccentricity
  increases with time until $e = 2\eforce$
  \citep{mori2004,scholl2007,meschiari2012,paarde2012,raf2013}. However,
  the particles are never actually stirred by the binary. They simply
  experience the independent free and forced modes of epicyclic
  oscillation, acting in concert but not in phase.

\item{\em Most-circular paths around a high-eccentricity binary.}  In Figure
  \ref{fig:kepecckep16ehixgrav}, the eccentricity of the Kepler-16
  binary is increased to $\ebin = 0.5$. The satellite is at 0.9~AU, just
  outside the critical radius for stable orbits for this choice of
  $\ebin$.  The Figure shows samples of the planet's position in the
  plane of the binary, taken over the course of $10^4$ satellite
  orbits (blue points in the Figure). The samples reveal the forced
  eccentricity and demonstrate that the periastron of the satellite is
  fixed and aligned with the binary. This numerical test is significant:
  the analytic theory described here is linear, but the $n$-body experiment 
  is not. The orbital alignment --- a prediction of the linear analytic
  theory --- holds in the non-linear case.

\item{\em Orbits in a massive disk}. Figure
  \ref{fig:kepecckep16ehixgrav} also demonstrates that most circular
  orbits do not precess even in the presence of a massive disk. We use
  a potential as in Equation~(\ref{eq:Sigma}) with a surface density
  of 2000~g/cm$^2$ at 1~AU. Around a single star with Kepler-16's mass,
  the disk causes rapid apsidal precession of a satellite at 0.9 AU, 
  $\dot{\varpi} \sim 0.005$~yr$^{-1}$ \citep{raf2013}.  However, around the 
  binary, the satellite's forced eccentric orbit is immune to precession 
  from the disk, as we expect from perturbation theory.  The example in 
  the Figure demonstrates that this prediction extends to moderately large 
  values of the binary's eccentricity.

\item{\em Gas drag.} Aside from changing the gravitational potential,
  gas modifies the orbits of small solids through aerodynamic drag.
  Figure~\ref{fig:kepeccdragkep16ehix} shows a particle orbit in a
  sub-Keplerian (``pressurized'') gas disk whose mean azimuthal speed
  is reduced by a factor of $1-\eta$ relative to a guiding center
  orbit in the disk's absence
  ($\eta = 0.001$; Equation~(\ref{eq:eta})). 
  For the purpose of this illustration, we do
  not include the disk's gravity. The drag force is proportional to
  the particle's speed in the local most circular reference frame of
  the gas. The magnitude of the force is sufficient to cause the
  particle to inspiral within a few hundred orbital periods. The
  Figure illustrates that even in the presence of gas drag, the
  particle orbits remain aligned with the binary.

\item{\em Effects of stirring: external time-dependent perturbations.}
  Collisional damping drives particles toward most circular orbits
  because particles coexist on these orbits without collisions. 
  Conversely, distant massive perturbers (i.e., planets) gravitationally 
  stir particles, driving them away from most circular paths. From the 
  analytic theory, we expect stirring to behave the same way around a binary 
  as it does around a single star. 
  Figure~\ref{fig:kepeccxkep16jupiter} shows simulation data from
  three scenarios: (i) a satellite like Kepler-16b;
  (ii) a satellite together with a Jupiter mass-body on a circular 
  orbit at 2~AU; and (iii) a satellite and a Jupiter orbiting a 
  central point mass.  Perturbations of the more distant planet affect 
  the motion of satellites in circumstellar and circumbinary cases 
  in much the same way, by generating nearly identical free eccentricity.

\item{\em Binary precession.} When an eccentric binary precesses, 
  we expect the satellite's forced eccentric orbit apsidally precesses 
  at the same rate.  Figure~\ref{fig:kepeccxkep16ehiwJx} illustrates 
  this behavior.  In a reference frame rotating with a precessing 
  eccentric ($\ebin = 0.5$) binary, the satellite's orbit shows no 
  precession due to a Jupiter mass perturber at 2~AU.

\end{itemize}

\subsection{Summary of the numerical studies}

To summarize, this set of numerical experiments illustrates that the 
main predictions of the linear analytic theory are confirmed in the 
non-linear regime.  A key result is that particles initialized on 
geometrically circular Keplerian orbits have a free eccentricity,
which leads to precession and possibly high-velocity collisions among
particles on adjacent orbits.  In contrast, most circular orbits remain 
nested and never cross, even in the presence of a massive disk or a gas 
giant perturber. Our conclusion is that the standard initial conditions
of particles on geometrically circular orbits are not realistic in a 
dynamically cool planetary disk. The nested most circular orbits are a 
better starting point. In the next section, we examine these conclusions 
in the context of planet formation theory.

\section{Circumbinary planet formation}

In the standard theory of star and planet formation, a rotating
molecular cloud of gas and dust collapses into a central protostar and
a circumstellar disk \citep[e.g.,][]{cass1981,tsc1984,yorke1993}. Disk
material is on nearly circular orbits \citep[e.g.,][]{weiden1977a};
viscous shear transports mass inward and angular momentum outward
\citep{lbp1974}.  Small solids are well-coupled to and flow with the
gas \citep[e.g.,][]{ada76,weiden1977a,raf2004,birn2010,chiang2010}.  
Larger solids decouple from the gas and follow Keplerian orbits 
about the central star.

Among the decoupled solids, various dynamical processes tend to
circularize their orbits around the central star.  Smaller particles
feel a strong headwind and are dragged towards the central
star. Although larger particles feel less drag, the gas efficiently
damps their orbits \citep[see
  also][]{saf1969,green1978,spaute1991,weth1993}. Along with gas drag,
collisional damping and dynamical friction drive particles towards
circular orbits \citep[e.g.,][]{horn1985,weth1993,kl1998,gold2004}.
As a result, most particles experience small relative collision
velocities which encourages growth through mergers \citep[see
  also][]{youdin2013}.

When the central protostar is a close binary, we expect a similar
evolution.  Infalling gas forms a circumbinary disk where the gas and
dust generally follow most circular paths around the binary.  Aside
from an inner gap in the disk roughly at $\acrit$, the structure of
the disk orbiting a circular binary is fairly similar to a
circumstellar disk around a single star
\citep[e.g.,][]{lin1979b,pri1991,arty1996}.  For eccentric binaries,
disk material appears to follow most circular orbits with forced
eccentricity, driven by the central binary
\citep[e.g.,][]{pierens2007,pel2013}.  As long as the disk
is not dominated by non-axisymmetric
structure (e.g., spiral density waves in a massive disk), we expect
pressure and viscosity to induce a smaller inward drift of the gas and
small particles relative to most circular orbits \citep[see
  also][]{pichardo2005,pichardo2008}. The gas attempts to circularize
the orbits of larger particles onto most circular orbits. Collisional
damping and dynamical friction also damp the orbits. Thus, large
particles end up on most circular orbits with a small amount of free
eccentricity with a magnitude similar to the eccentricity of particles
in disks around a single star.

Achieving these configurations is a natural long-term outcome for
dissipative disks. However, it is worth considering how quickly the
gas, dust, and larger solid particles circularize.  We then
compare these time scales with time scales for other processes such as
precession of the binary. We expect that particles settle onto most 
circular orbits when the precession time scales are long relative
to circularization time scales.  Here, we consider several basic time
scales, treating gas-dominated and particle disks as separate cases
due to differences in their surface densities and in the physical
processes that drive them.

\subsection{Gas disks}

We start by considering a massive circumbinary disk to model
primordial gas in protoplanetary systems
\citep{weiden1977b,hayashi1981}.  As before, we choose a
surface density
\begin{eqnarray}
\Sigma_g(a) &  = & \Sigma_{g,0} \left(\frac{a}{a_0}\right)^{-1}
\end{eqnarray}
where $\Sigma_{g,0} = 2000$~g/cm$^2$, and $a_0 = 1$~AU.
The disk has a vertical scale height 
\begin{equation}
h(a) = h_{g,0}\left(\frac{a}{a_0}\right)^q
\end{equation}
where $h_{g,0}/a_0 \sim 0.02$ and $q = 9/7$ \citep{chiang1997}.

Damping times in this disk derive from the sound speed,
\begin{equation}
c_s(a) \sim h(a)\vkep/a, 
\end{equation}
where $\vkep$ is the orbital velocity at distance $a$.  The time scale
for vertical structures to reach hydrostatic equilibrium is $h/c_s$, 
close to the dynamical time \citep{lightman1974,pri1981}.
Pressure damping of features along an orbital path require at least
\begin{eqnarray}\label{eq:tdampgas}
\tgdamp & \gtrsim &  a/c_s \sim \frac{a}{h}\tkep  
\approx 8 \aAU^{17/14} \MMsun^{-1/2} \textrm{yr}.
\end{eqnarray}
where $M$ is the total central mass. Radial structures may dissipate
more slowly, on a diffusive time scale 
\citep[e.g.,][]{ogilvie2001}
\begin{eqnarray}\label{eq:tdampgasII}
\tgdamp & \lesssim &  \frac{a^2}{h^2}\tkep  
\approx 2.5\times 10^3 \aAU^{13/14} \MMsun^{-1/2} \textrm{yr}.
\end{eqnarray}
We anticipate that disks with free eccentricity damp to most
circular orbits on time scales in this range.

By comparison, the time scale for binary precession, a result
of the gravitational interaction between the gas disk and the binary, 
is long. If the inner edge of the disk is at orbital distance $\ain$,
then the binary's apsidal precession rate is
\cite[e.g.,][]{ward1981,raf2013}:
\begin{eqnarray}
\apsprebin & \approx & 0.5 \frac{\pi G \Sigma_g(\ain)}{\nbin\ain}
              \,\left(\frac{\abin}{\ain}\right) ~ .
\end{eqnarray}
The precession time is then
\begin{equation}\label{eq:tbinpregas}
\tprecbin \approx 2.0\times 10^4 \SigmagOhMMcgs^{-1}\abinOhpIIAU^{1/2}
\ainIIpVbin^3\ \textrm{yr} \ \ \ \ \textrm{[gas disk]}, 
\end{equation}
where we choose to set $\ain$ to be the innermost stable orbit at
$\acrit \approx 2.5\,\abin$, for the values of the parameters in the
angular brackets.  

Thus, binary precession is slow compared with dynamical times. 
Precession is a perturbative effect. We therefore expect that orbits in
the disk remain apsidally aligned with the binary.  We conclude that 
gas disks damp quickly to most circular orbits, so long as the physics 
(gravity, hydrodynamics) enables the disk to be axisymmetric when 
averaged over these orbits \citep[although see][]{pel2013}. 

\subsection{Particle disks}

As the gaseous disk evolves, solids particles evolve with it. Small particles
remain coupled to the gas. Larger particles with sizes of 1~cm or larger
(depending on the local properties of the gas) are uncoupled 
\citep[e.g.,][]{ada76,weiden1977a,raf2004,birn2010}. Collisional processes
enable some particles to grow to larger sizes 
\citep[e.g.,][]{windmark2012,garaud2013}.

For large particles, interactions with the gas produce a radial drift and 
circularize the orbits. The characteristic time scale for these interactions is
\citep{ada76,weiden1977a,chiang2010}:
\begin{equation}
\label{eq:tdrift}
\tcirc \approx 15
  \left ( {\rp \over {\rm 1~km} } \right )
  \left ( { {\rm 1~AU} \over a} \right ) 
  \left ( { 10^{-9}~\rm g~cm^{-3} \over \rho_g} \right ) \tkep,
\end{equation}
where $\rp$ is the radius of the particle, $\rho_g$ is the local gas volume
density (normalized to the typical density of a minimum mass solar
nebula), and $\tkep$ is the orbital period of the local guiding center.
For all but very large objects with $\rp \gtrsim$ 100~km, this time scale 
is small compared with the precession time. Thus, small particles uncoupled
from the gas likely find most circular orbits.

Numerical simulations of ensembles of small particles interacting with
the gas suggest the surface density of the solids is somewhat steeper 
than the surface density of the gas 
\citep[e.g.,][]{youdin2004a,brauer2008,birn2010,birn2012,laibe2012,pinte2014}.
Here we adopt a standard power law:
\begin{eqnarray}
\Sigma(a) &  = & \Sigma_{0} \left(\frac{a}{a_0}\right)^{-1.5}
\end{eqnarray}
where $\Sigma_{0} = 10$~g/cm$^2$. To derive basic time scales,
we assume that particles in the disk have radii of $\rp = 1$~km 
and a density of 3~g/cm$^3$. If they are stirred to their escape 
velocity,
\begin{eqnarray}
\vesc & = & \sqrt{2Gm/r} = \sqrt{(8\pi G\rho/ 3)} \, r
\\  \nonumber
\ & = & 130\rhoIIIcgs^{1/2}\rkm \ \textrm{cm/s}, 
\end{eqnarray}
then we can use standard kinetic theory to estimate the collision time
as $(n \sigma v)^{-1}$ where $n$ is the number density, $\sigma$ is the 
collisional cross-section, and $v$ is the relative velocity. The damping 
time from collision is then
\begin{eqnarray}
\label{eq:tdamprkm}
\tdamp & \sim & \frac{\rho \rp\tkep }{2\sqrt{2}\pi\Sigma}
\\ \nonumber
\ &  \approx &
3400 \rhoIIIcgs \rkm \aAU^{6} \SigmaOhXcgs^{-1} 
      \MMsun^{-1/2} \textrm{yr}.
\end{eqnarray}
where we assume that collisions are inelastic and the disk scale height 
is proportional to the relative speed $v$, divided by the orbital frequency. 
This expression ignores gravitational focusing, which reduces the 
damping time \citep[e.g.,][]{weth1993,oht1999}.

While the damping time is only linearly dependent on particle size,
the range of sizes can be considerable in an evolving planet-forming
disk. For pebbles ($\rp \approx 1$~cm), the damping time is very fast:
$\tdamp$ is formally shorter than a dynamical time.  For large
particles ($\gae 1$~km), the time scale for collisional damping is
fairly long, approaching the precession time scale for the gas when
$\rp \gtrsim$ 10~km. However, these larger particles interact with 
smaller particles through dynamical friction and viscous stirring
\citep[e.g.,][]{oht2002,gold2004}. For these processes, the typical
damping time for large particles is a factor of 10--100 smaller than
suggested by Equation (\ref{eq:tdamprkm}). Thus, damping times for
particles with $\rp \lesssim$ 100~km are short compared with the precession
time.

For comparison, the gravity of the particle ring
induces the binary to precess with a period
\begin{equation}\label{eq:tbinpreparticles}
\tprecbin \approx 4.0\times 10^6 \SigmaOhXcgs^{-1}\abinOhpIIAU^{1/2}
\ainIIpVbin^3\ \textrm{yr}  \ \ \ \ \textrm{[particle disk]}.
\end{equation}
We conclude that in a particle disk, as in a gas disk, conditions allow for
damping to most circular orbits, so they may serve as the equivalent
to circular Keplerian orbits. 

Eventually, the largest particles contain more than half the mass of 
all the solids in an annulus of the disk. At this point, damping times 
become much longer than stirring times; orbital evolution then 
becomes chaotic \citep[e.g.,][]{chambers1998,gold2004,kb2006}. Around
a single star, chaotic systems eventually settle down into stable 
multi-planet systems. Aside from the impact of the inner unstable region
and resonances, we expect a similar evolution of chaotic systems around
binary stars \citep[e.g.,][]{quintana2006}.

\subsection{Perturbations from massive planets}

In a particle disk, a circumbinary Jupiter-mass planet dramatically 
shortens the precession time for the binary.  If planetesimals orbit 
within a few times the binary separation and the massive planet orbits 
well outside this region (e.g., as in
Figure~\ref{fig:kepeccxkep16jupiter}, $\abin\sim 0.2$~AU, satellite at
$a \sim 1$~AU and ``jupiter'' at 2~AU), then the binary precession
rate is
\begin{equation}\label{eq:precessjup}
\apsprebin \approx \frac{3}{4}\frac{m_{\rm j}}{M} 
\frac{\abin^3}{a_{\rm j}^3} \nbin, 
\end{equation}
where $M_{\rm j}$ and $a_{\rm j}$ are the planet's mass and orbital
distance.  Substituting parameter values comparable to observed binary
systems and consistent with the set-up in 
Figure~\ref{fig:kepeccxkep16jupiter}, the precession period
($2\pi/\apsprebin$) is
\begin{equation}\label{eq:tbinpregasgiant}
\tprecbin \approx 1.2\times 10^5 \MMsun^{1/2} \mjMJ^{-1}
\abinOhpIIAU^{-1/2}\ajIIAU^2 \ \textrm{yr}.
\end{equation}
Any binary precession from the gas giant perturber is a small perturbative 
effect that preserves the apsidal alignment between binary and planetesimal 
orbits.

\subsection{Planet formation in circumbinary disks}

Our analysis establishes time scales for gas and particles to damp to
most circular orbits.  With no precession of the binary, these orbits
do not precess. Although binary precession induces precession in most
circular orbits, these orbits remained apsidally aligned with the binary.
In this way, most circular orbits provide reference frames in which the 
dynamics of planet formation takes place 
\citep[see also][]{pichardo2005,pichardo2008}.  For example, collisional
damping and self-stirring of planetesimals modifies the free eccentricity
of circumbinary planetesimals \citep[e.g., as in][]{weth1980}; these 
processes have no impact on the forced eccentricity (and high-frequency 
modes) driven by the central binary.  
Gravitational stirring from distant 
planets \citep[e.g.,][]{weiden1989} also excites eccentricity,
driving epicyclic motion in a manner similar to the binary itself 
\citep[see Figure~\ref{fig:kepeccxkep16jupiter}]{hepp1978}.

When ensembles of orbiting planetesimals attempt to grow into a planetary
system, the relative velocity between particles,
\begin{equation}
v \sim \efree\vkep,
\end{equation}
is a key parameter which establishes the efficiency of gravitational
focusing and collision outcomes \citep{stewart1988,weth1993,oht1999,
oht2002}. Small
relative velocities favor growth by mergers; large relative velocities
favor destruction. To estimate the boundary between these regimes, we
rely on the specific collision energy $Q_d^\ast$ required to disperse
half the mass of a colliding pair of planetesimals to infinity. For
rocky material,
\begin{equation}
Q_d^\ast \approx 3\times 10^5 \rkm^{-0.4} + 7 \times 10^6 \rhoIIIcgs \rkm^{1.35}
\ \textrm{erg/g}
\end{equation}
\citep{davis1985,weth1993,benz1999,housen1999,kb2005}. Equating $Q_d^\ast$
to the specific rest-frame collision energy between two equal-mass objects, 
mass loss exceeds mass gain when
\begin{equation}
\vdestruct \gae 0.1 \ \textrm{km/s \ \ \ \ \ [destructive collisions, $r=1$\,km]}
\end{equation}
for kilometer-size rocky planetesimals. 
This relative speed is best measured in the reference frames of most circular 
orbits.

Around a binary at orbital distance $a$, all planetesimals have an
additional component to their velocity compared with their counterparts
around single stars.  This velocity, from the forced eccentricity, has a
typical magnitude
\begin{eqnarray}
v \sim \vforce & \equiv & \eforce\vkep \approx \frac{(\Mp-\Ms)}{M}
\frac{\abin}{a}\ebin\vkep \\
\ & \approx & 
0.72 \dMhalfMsun \MMsun^{-1/2}\abinOhpIIAU \ebinOhpII \aAU^{-3/2}
\ \textrm{km/s} ~ .
\end{eqnarray}
Comparing this value with the disruption speed, it is clear that
interpreting $\vforce$ as a random motion leads to the simple 
prediction of destructive collisions for a broad range of particle
sizes. Yet $\vforce$ is the speed of a reference frame tied to a most 
circular orbit. Particles traveling on most circular orbits have no 
relative radial velocity;  planetesimals may perturb each other and collide,
 but at velocities much smaller than 
$\vdestruct$ and $\vforce$, promoting mergers instead of destruction.

This picture differs from the approach often taken in studies of
circumbinary planet formation. For example, in \citet{mori2004},
\citet{meschiari2012}, \citet{paarde2012}
and \citet{lines2014}, particles in $n$-body simulations are
initialized on circular Keplerian orbits about the binary center of
mass. Gas, if present, is assumed to have fluid elements on exactly circular 
sub-Keplerian orbits. 
From \S\ref{sec:theory}
a particle trajectory
with total eccentricity of $e=0$ is identical to a trajectory with
equal parts free and forced eccentricity ($\efree=\eforce$), where the
phase of the free part is chosen to yield an initial net eccentricity
of zero ($\psi_e = \pi$).  
The difficulty with these initial
conditions is that the relative velocities are set with $v \sim
\efree\vkep \sim \vforce$.  It is then just a matter of time before
precession of the free epicyclic motion drifts from the force motion,
and particle orbits can cross.  That time is
\begin{equation}
\tpreaps = \frac{2\pi}{\dot{\varpi}} \approx 67 \abinOhpIIAU^{-2}
\aAU^{7/2} \MMsun^{1/2} \muhalfMsun^{-1} \ \textrm{yr,}
\end{equation}
where the reduced mass is $\mreduced<0.5 M$. 
Entrainment of solids by the gas on sub-Keplerian orbits can reduce relative
velocities, but only for particles of similar size
\citep[e.g.,][]{marzari2000}.
Thus, setting up protoplanetary disks with all objects having comparable 
free and forced eccentricities quickly dooms them to 
destruction.

Sometimes, the initial orbits of circumbinary planetesimals have no 
impact on the outcome of planet formation calculations.  For example, 
if the free eccentricity is much larger than the forced eccentricity, 
then the binary's time varying potential is an unnoticed perturbation.  
This situation applies in the late stages of planet formation, when 
most of the mass is concentrated into a small number of large objects. 
At this stage, large objects have random velocities comparable to the
escape velocity of the largest object \citep[e.g.,][]{gold2004,kb2006}.
Setting $\vesc \gae \vforce$, we obtain
\begin{equation}
r_{\rm pro} \gtrsim 1000 \rhoIIIcgs^{-1/2}\MMsun^{1/2}\dMhalfMsun
\abinOhpIIAU\ebinOhpII\aAU^{-3/2}\ \textrm{km}.
\end{equation}
Outside of the critical radius for stability, $\acrit$, a simulation 
with a collection of Pluto-size or larger objects should fail to see 
a difference between circumstellar and circumbinary environments.
The simulations of \citet{quintana2006} support this interpretation.
Except for the effects of the destabilization of orbits near $\acrit$, 
ensembles of roughly lunar mass objects grow into a few terrestrial 
planets in circumstellar and circumbinary environments.

\subsection{Additional issues: resonances and binary evolution}

Once planetesimals have settled around most circular paths to grow by
coagulation, the binary still influences their evolution.  Overlapping
resonances may stir particles and destabilize them \citep{wisdom1980}.
However, these unstable orbits are typically close to the binary. For
example, the 3:1 resonance, which corresponds to a singularity in the
denominator of Equation~(\ref{eq:Cpmk}), lies within $2.1\abin$.  As a
guide, these unstable orbits are usually within the critical radius
$\acrit$ identified by \citet{holman1999}.  In some binary
configurations instabilities extend beyond $\acrit$
\citep[e.g.,][]{musielak2005,pichardo2005,pichardo2008,doolin2011}.
Numerical simulations provide a straightforward way to identify these
orbits \citep[e.g.][]{popova2013,chavez2015} and to understand
any impact on gas or solid particles in the planetary disk
\citep[e.g.,][]{pierens2008b}.

Tidal evolution of the binary also complicates the long-term
stability of circumbinary planetary systems. Changes in the orbital
separation and eccentricity can change the positions of critical
resonances, leading to instability in systems which had been
stable. In the Pluto-Charon system, satellites become unstable when
tidal evolution drives a relatively rapid expansion of the binary
orbit \citep{ward2006,lith2008a,cheng2014b}. For binaries composed of
solar-type stars, tidal evolution on time scales of 1--100~Myr is
important only for systems with orbital periods of 10 days or less
\citep{meibom2005}.  Thus, tidal evolution of the central binary
probably has little impact on the formation of most circumbinary
planetary systems.

\subsection{Summary}

Unless the collapse of a molecular cloud into a binary + disk system
is significantly different from the collapse to a single star + disk 
system, we expect the stages of circumbinary planet formation to parallel 
those of circumstellar planet formation. As long as non-axisymmetric
structures such as spiral density waves are relatively unimportant,
the time scales for gas damping, gas drag, collisional damping, and 
dynamical friction are all much shorter than typical precession times 
for the binary. Thus, material in a circumbinary disk should find its 
way onto most circular orbits around the binary in a similar way as 
material in a circumstellar disk damps onto circular orbits around a 
single star. Once material lies on or close to most circular orbits, 
small relative velocities of particles on adjacent orbits strongly 
favors growth over collisional disruption.

Precession of the inner binary has little impact on this conclusion.
When a massive disk or a giant planet causes the inner binary to precess,
the orbits of solid particles on most circular orbits maintain apsidal
alignment. 
Because relevant damping time scales are short ($\lesssim 10$~yr for
gas and $\lesssim 10^3$~yr for particles) compared with the binary
precession time scales ($\gtrsim 10^4$~yr for precession induced by a
massive disk, or longer with particle disks or gas giants beyond 1~AU)
both gas and solids can maintain this apsidal alignment (see Equations
(\ref{eq:tdampgas}) and (\ref{eq:tdamprkm}) for damping times and
Equations (\ref{eq:tbinpregas}), (\ref{eq:tbinpreparticles}) and
(\ref{eq:tbinpregasgiant}) for binary precession periods).
Most circular orbits remain the reference frame in which to measure
relative velocities. Thus, planet formation proceeds in a standard
fashion.

For calculations of circumbinary planet formation, starting with the
right initial conditions for orbits of gaseous material or solid 
particles is crucial. Around a single star, it is sufficient to start
material on orbits with modest eccentricity: precession is relatively
unimportant and all damping processes lead to circular orbits. Around
a binary, it is important to differentiate between the forced
eccentricity driven by the binary and the free eccentricity relative
to a most circular orbit around the binary 
\citep[see also][]{pichardo2005,pichardo2008}. Starting particles on
circular Keplerian orbits around the binary center of mass creates a
free eccentricity relative to a most circular orbit. In general, this
free eccentricity is comparable to the forced eccentricity induced by
the binary. For eccentric binaries, this extra eccentricity produces
spuriously large collision velocities which can lead to collisional
destruction instead of growth by merger. To provide a proper 
evaluation of circumbinary planet formation, it is essential to begin
with most circular orbits and then evaluate relative collision 
velocities for appropriate  values of the free eccentricity.

\section{Application: The \textit{Kepler} circumbinary planets}
\label{sec:kepler}

In the last decade, data from the \kepler\ satellite have started to 
paint a rich picture of circumbinary planetary systems 
\citep[e.g.,][and references therein]{armstrong2014}. The known systems
\citep{doyle2011,welsh2012,orosz2012a,orosz2012b,schwamb2013, kostov2013,
kostov2014}
have fairly massive planets with radii $\rp \approx$ 0.25--0.76 $R_J$ and 
semimajor axes $a \approx$ 0.3--1.1~AU orbiting binaries with a broad 
range of mass ratios (0.25--1) and eccentricities (0.02--0.52).  
Although the number of circumbinary planetary systems is still rather 
small ($\sim$ 10), detection rates suggest these planets are roughly 
as common as planets around single stars \citep{armstrong2014}. 

At larger semimajor axes, circumbinary debris disks are also common.
Among main sequence stars with FGK spectral types, single stars and
binaries are equally likely to have debris disks \citep{trill2007}.
In circumbinary debris disks, the disk and binary are often co-planar
\citep[e.g.,][]{kennedy2012,kennedy2015}. 
Small number statistics currently prevents robust conclusions, but it 
seems plausible that this structure is primordial: circumbinary disks
form in the plane of the binary system \citep{kennedy2012,kennedy2015}.

Understanding the structure of primordial circumbinary disks is also
hampered by small sample sizes \citep[e.g.,][]{harris2012}. Among
multiple stars with ages of 1--3~Myr in the Taurus-Auriga molecular
cloud, disks around single stars have masses similar to those of very
wide binaries with $a \gtrsim$ 300~AU (0.001--0.1~\Msolar, on the basis
of dust mass estimates from millimeter fluxes; \citealt{and2005}).
Binaries with $a \approx$ 30--300~AU (5--30~AU) have factor of 3--5
(5--10) smaller disk masses. In three binaries with separations
ranging from about 0.05~AU to 10~AU, disk mass estimates yield
0.01--0.03~\Msolar, comparable to the disks in single stars and wide
binaries. For reference, a mass of 0.01~\Msolar\ corresponds to the
Minimum-Mass Solar Nebula (MMSN), which is a the lower limit on the
mass required to build the planets in the solar system
\citep{weiden1977b,hayashi1981}.

Here, we survey known \kepler\ circumbinary systems to interpret how 
their orbital characteristics and size might inform us of their origin. 
While we reiterate some of the discussion in \citet{leung2013}, we also 
focus on how the observations might impact our understanding of planet 
formation scenarios, whether these planets formed {\it in situ} or 
migrated from some larger semimajor axis inward.

To frame the problem, we consider several different formation scenarios:
\begin{enumerate}[I.]
\item {\it{\it In situ} formation with no migration.} Planets grow by
  coagulation from nearby gas and dust. The final planetary mass is
  limited by the {\it initial} surface density of the protoplanetary
  disk.  A Minimum-Mass Solar Nebula with $\Sigma_0 = 7$~g/cm$^2$
  in Equation~(\ref{eq:Sigma}) has 4~\Mearth\ in solids inside of the
  snow line a 2.7~AU and roughly 50~\Mearth\ in solids inside 50~AU
  \citep{weiden1977b,hayashi1981}.  The mass in gas is roughly
  800~\Mearth\ inside 2.7~AU and roughly 3000~\mearth\ inside 50~AU.
  Numerical simulations suggest rocky planet formation is inefficient;
  collisional processes often lose significant amounts of mass during
  the assembly of Earth-mass and larger planets
  \citep[e.g.,][]{kb2004b,kb2006,raymond2006,raymond2011}. Augmenting
  the MMSN by a factor of 2--3 is sufficient to yield Earth-mass
  planets at 0.7--1 AU in the solar system.

  Producing a Neptune-mass planet {\it in situ} at 1 AU is more
  challenging.  Accreting sufficient gas from the disk to produce
  Neptune requires a $\sim$ 10~\mearth\ solid core
  \citep[e.g.,][]{pollack1996,raf2011,rogers2011,piso2015}.  In the 
  standard MMSN, the mass in solids is insufficient to produce such
  a massive core. When the surface density of the disk is roughly 
  20 times the MMSN, coagulation models routinely yield 
  10~\mearth\ cores \citep[e.g.,][see below]{hansen2012,schlichting2014}.

  In most numerical simulations, formation of 2--3 Earth mass or
  larger planets is common; producing a single planet is rare
  \citep[e.g.,][]{kb2006,raymond2011,hansen2012,hansen2014b}.  If
  these simulations are `missing' an important piece of physics which
  allows several Earth-mass planets to merge into a single super-Earth
  mass planet, {\it in situ} formation of single super-Earth or
  Neptune mass planets might be possible in lower mass disks, e.g.,
  3--5 times the mass of the MMSN instead of $\sim$ 20 times. Without
  a better understanding, we assume that formation of Neptune mass
  planets requires a very massive disk.

\item{\it Migration then assembly.} Precursor solid material is first moved 
  from large $a$ to within 1~AU of the host star. Planet formation then
  proceeds {\it in situ} \citep{hansen2012}. In this way, the limited
  amount of solids available inside of 1~AU is enhanced at early times;
  the delivery mechanism is uncertain.  Because of the structure of the 
  protoplanetary disk at small radii, gas accretion may be less efficient 
  than beyond the snow line. Planets formed in this way may be ``gas-starved'' 
  compared with more distant gas giants.

\item{\it Migration through a gas disk.}  Gas giants form beyond the snow 
  line, where the solid-to-gas ratio is a factor of $\sim$ 3 larger
  \citep[e.g.,][]{kk2008}. Planets move inward by exchanging torque 
  with the gas disk \citep{ward1997}. Composition and structure of these 
  planets are typical of the solar system's gas giants.

\item{\it Planet-planet scattering.}  Massive planets formed beyond
  the snow line gravitationally scatter one another, either inward
  toward the central mass or outward
  \citep{rasio1996,chatterjee2008,juric2008,raymond2010,marzari2010,
  beauge2012,moeckel2012,bk2014}. In a scattering event
  between two planets, if the more massive body is Neptune-size or
  larger (assuming typical solar system densities and orbital
  parameters for gas giants), then a broad range of outcomes is
  possible, including ejection from the system or the placement of a 
  ``hot Jupiter'' near a single central star.  High orbital 
  eccentricity is the red flag for these scattering events.
\end{enumerate}

For each of these modes, the initial mass of the disk and the epoch
of planet formation are important considerations 
\citep[e.g.,][and references therein]{najita2014}. In the Taurus-Auriga 
molecular cloud \citep[e.g.,][]{and2005,and2013}, the median mass in 
solids for disks around single stars drops from 50--100~\mearth\ for 
protostars (ages of 0.1--0.5~Myr) to 10--20~\mearth\ for T Tauri stars 
(ages of 1--3~Myr). Observations of other star-forming regions suggest
this evolution is typical \citep[e.g.,][]{and2007,will2011}. Thus, 
existence of Neptune-mass to Jupiter mass planets favors early formation
in the massive disks of young protostars\footnote{For the solar system, 
radiometric analyses of meteorites similarly suggests formation of solids
when the Sun had an age of 0.1--0.3~Myr \citep[e.g.,][]{kleine2009,dauphas2011a}.}.

\subsection{Disk mass requirements for \textit{in situ} growth}

For {\it in situ} formation, we estimate the surface density required 
to make a planet with mass $m$ and radius $r$.  Each protoplanet 
accretes material from a ``feeding zone'' with radial width $\delta a$. 
Random motions of particles near the protoplanet set this width. When
these particles have relative speeds smaller than the planet's escape 
velocity, the planet can accrete them. For particles with a free eccentricity, 
$e_{free} \lesssim \vesc/\vkep$ \citep[e.g.,][]{schlichting2014}. Thus,
\begin{equation}\label{eq:feedzone}
\Delta a \sim 2 a \efree \sim 2 a\frac{\vesc}{\vkep} \sim 2 a \left(\frac{2 a m}{r M}\right)^{1/2}.
\end{equation}
where $m$ and $r$ are the core's mass and radius\footnote{This result
for the width of the feeding zone is somewhat larger and probably more 
realistic than the width derived for the more standard ``isolation mass'' 
\citep{green1990}.}.  By choosing a form for the surface density
(we use $\Sigma \sim a^{-1.5}$ as in Equation~(\ref{eq:Sigma})) and
integrating over the feeding zone (excluding any unstable region
inside of $\acrit$) we evaluate the minimum value for $\Sigma$ at
the planet's position.

Figure \ref{fig:kepsigma} shows the results of applying this prescription 
to the \kepler\ circumbinary planets in Table~\ref{tab:kepler}. 
The minimum surface density required to build gas-accreting cores is 
roughly an order of magnitude more than the MMSN. Thus, any {\it in situ} 
formation model requires a massive disk. Models which form these planets
well outside the snow line require much less massive disks. 

This result allows us to eliminate {\it in situ} (no migration) models
for all of the \kepler\ circumbinary planets.  If the \kepler\ circumbinary
planets are common \citep{armstrong2014}, the precursor disks are also 
common.  Among the youngests stars with ages of 0.1--0.5~Myr, however,
disks with initial surface densities of 10--20 times the MMSN are 
exceedingly rare \citep{and2005}. Because lower mass protostellar disks
are common, the \kepler\ planets require a formation model with some form 
of migration or scattering.

\subsection{Resonances, instabilities and migration}

Although formation followed by radial migration is a plausible path
for the \kepler\ circumbinary planets, unstable orbital resonances in
the circumbinary environment pose clear obstacles to migration
\citep{pierens2007}. Near certain commensurabilities, overlapping
resonances excite large eccentricities \citep{wisdom:1982,
  wisdom:1983, lecar2001}.  The \citet{holman1999} condition that
orbits are unstable if they are within a distance of $\acrit$ from the
binary center of mass is a working guide; refinements are needed to
identify the presence and ``strength'' of unstable resonances even
outside of $\acrit$ \citep[e.g.,][]{popova2013,chavez2015}.

Here, we assess whether the region around each binary in
Table~\ref{tab:kepler} is stable as a planet migrates inward.  We
consider two approaches: directly migrating the planet radially inward
and, equivalently, expanding the binary. We show results for binary
expansion models, which are more straightforward (and stringent) since
they require adjusting only the Keplerian semimajor axis of the
binary, instead of modifying the non-Keplerian orbit of the planet to
mimic radial drift.  Figure~\ref{fig:kepeccrez} shows the results for
each \kepler\ planet at an equivalent radial drift rate of less than
$10^{-5}$~AU/yr \citep[typical for planets undergoing type I
  migration;][]{gold1980,ward1997,tanaka2002} The particles in the
Figure manage to get as close to their binary host as the planet's
current position. Once inside, they become unstable near the 5:1
resonance, except for Kepler-34b, which becomes unstable near the 7:1
commensurability. Kepler-47b, orbiting the binary with the lowest
eccentricity, is stable down to the 4:1 resonance. Thus, it seems
plausible that the \kepler\ circumbinary planets, or smaller
precursors (with fast damping times), could migrate through the
potential minefield of resonances.  Further tests would be needed to
confirm whether the stability that we observe would remain with lower
damping rates or slower migration times.

\subsection{The \kepler\ circumbinary planets}

With constraints on {\it in situ} formation and migration from
large distances established, we consider whether any of the four 
modes of planet formation is consistent with the bulk properties
and orbit of each \kepler\ circumbinary planet listed in 
Table~\ref{tab:kepler}.  We start with the first and most famous 
discovery, Kepler-16b.

\begin{itemize}
\item{\bf Kepler-16.} The central binary, with a K-dwarf primary and
  an M-dwarf secondary, has significant eccentricity ($\ebin \approx
  0.16$). The planet is a gas giant with radius of 0.7~$R_J$ and a
  mass of 0.3~$M_J$, orbiting at 0.7~AU with low eccentricity, 
  $\sim 0.01$ \citep{doyle2011}. From the analytical theory 
  (\S\ref{sec:theory}), \citet{leung2013} estimate that free and 
  forced eccentricities are comparable, $\efree \approx
  \eforce \approx 0.03$. 

  The high mass and low eccentricity of the planet favor the 
  migrate-then-assemble and the migrate-in-gas modes of planet
  formation.  {\it In situ} formation with no migration, although 
  plausible, requires a very massive disk which is very uncommon
  among low mass stars with ages of 0.1--0.5~Myr 
  (see Figure~\ref{fig:kepsigma}).
  Planet-planet scattering is less likely to result in
  this low-eccentricity configuration
  \citep[e.g.,][]{marzari2010,beauge2012}.

\item{\bf Kepler-34.} The binary has high eccentricity ($\ebin \approx
  0.5$) but the mass difference between the partners is small, within
  a few percent.  Thus, the forced eccentricity at Kepler-34b's
  location is low, around 0.002. In contrast, the eccentricity of the
  planet is much larger, exceeding 0.2. The planet is about the same size
  as Kepler-16b and may be a Saturn analog.  The high eccentricity
  and high mass favor a scattering scenario. 

\item{\bf Kepler-35.} This binary has stars that are also nearly equal
  in mass. The planet, Kepler-35b, is nearly the same size as Kepler-16b 
  and Kepler-34b, but has a much smaller eccentricity (0.048). As with 
  Kepler-16b, the migrate-then-assemble or migrate-in-gas modes are 
  strongly favored over the scattering or {\it in situ} with no migration
  formation modes.

\item{\bf Kepler-38.} The mass ratio of the binary is almost five to
  one; the eccentricity is around 0.1. The planet is comparatively small, 
  with a radius of 0.4 $R_J$, suggesting a mass of roughly 20~\Mearth. 
  Despite its low eccentricity (consistent with zero) and low mass, 
  {\it in situ} formation with no migration is still problematic. 
  Kepler-38b requires a surface density that is more than a factor of 
  twenty larger than the MMSN. The low eccentricity similarly eliminates
  the scattering mode. Thus, we favor either the migrate-then-assemble or 
  migrate-in-gas mode.

\item{\bf Kepler-47.} The binary consists of a Sun-like star and a red dwarf 
  on a nearly circular orbit. The forced eccentricity of its planet is low; 
  the eccentricity of the planet is consistent with zero.  With a radius just 
  under 3~R$_\oplus$, Kepler-47b probably has the smallest mass of all the 
  known circumbinary planets, perhaps 10~\Mearth.  This planet has a companion 
  about twice its radius, Kepler-47c, at an orbital distance of about 1~AU. 
  The combined mass of the two planets works against {\it in situ} formation 
  with no migration. Two planets orbiting inside the snow line requires a
  complicated scattering scenario \citep[see][]{moeckel2012}. The
  migrate-then-assemble and the migrate-in-gas modes are less complicated.
  Improved statistics for this kind of circumbinary planetary system would
  provide better constraints on the scattering mode.

\item{\bf PH1/Kepler-64.} This binary has both a significant mass
  difference between its partners and a modestly high eccentricity
  (0.2).  The resulting forced eccentricity is the largest of this
  sample (0.044). The planet PH1b is half the size of Jupiter, with an 
  eccentricity of 0.05; thus, the free eccentricity is probably low.  
  As with the other \kepler\ circumbinary planets, the large mass of the
  planet precludes {\it in situ} formation with no migration; the small
  orbital eccentricity eliminates most scattering models. Early formation
  in a massive disk allows either the migrate-then-assemble or the
  migrate-in-gas models.

\item{\bf Kepler-413.} This low-eccentricity binary hosts a modest
  size planet, a third the radius of Jupiter. Like Kepler-38b, this
  mass is uncomfortably high for {\it in situ} formation without
  migration. Unlike Kepler-38b, Kepler-413b has a significant free 
  eccentricity of more than 0.1.  Large eccentricity tends to preclude
  the migrate-in-gas mode. However, either the migrate-then-assemble 
  mode or a scattering model can produce a planet similar to Kepler-413b.

\end{itemize}

\subsection{Summary}

The circumbinary planets observed by \kepler\ are Neptune-size or bigger 
and located just beyond the critical radius around their host stars. 
Although {\it in situ} planet formation with no migration is a 
promising way to grow Earth mass planets at these distances, it is
very unlikely to be responsible for the known \kepler\ circumbinary 
planets (Figure~\ref{fig:kepsigma}).  The central issue --- not enough 
mass to build Neptune-size planets --- is circumvented by importing solids
from beyond the snow line. Other formation mechanisms accomplish this 
mass transfer by invoking an inward radial flux of small particles within 
the disk, migrating fully-formed gas giants through the disk, or 
scattering gas giants from outside the snow-line where other large 
planets form.

Migration seems to be involved in most of the \kepler\ planets.
However, without larger
samples of planets, it is impossible to distinguish between models where
migration precedes assembly from those where migration follows assembly.
All of the planets are too massive to allow {\it in situ} formation with
no migration. However, the high free eccentricity observed in Kepler-34b
and Kepler-413b are consistent with scattering events. Improved constraints
on the orbits and bulk properties (mass, composition, spin, etc.) might
allow more rigorous conclusions on their origin.

Here our most important contribution to the discussion of close-in
planets \citep[e.g.,][]{schlichting2014} is not to discriminate between 
formation mechanisms, but to emphasize that all are viable in the 
circumbinary environment at $a \gtrsim \acrit$. Thus, all issues
concerning the formation of Neptunes and Jupiters inside the snow 
line for a single star carry over to the circumbinary case.

\section{Discussion}\label{sec:discussion}

Our main conclusion is that outside of a small region near a binary 
star, planet formation proceeds in much the same way as around a 
single star. This result stems from the existence of a family of
``most circular'' orbits around binaries that do not intersect,
analogous to concentric circles around a single star.  
Gas, dust, and growing planets orbitally damp to these streamlined
paths to avoid mutual collisions. The growth of planets, involving
mergers, fragmentation, stirring and dynamical friction, takes place
in the reference frame of guiding paths on these most circular orbits,
just as it does in circumstellar disks. Without such paths,
planetesimal orbits would inevitably mix at high velocity, leading to
destruction, not growth.

These most circular paths are rooted in analytical theory.
\citet{lee2006} and \citet{leung2013} lay out the foundation,
describing how circumbinary orbits are approximated as linear
combinations of (i) rapid, forced oscillations in response to the
time-varying potential, (ii) slower, epicyclic motion --- the ``forced
eccentricity'' --- that responds to the binary's eccentricity, and
(iii) ``free eccentricity'' and inclination relative to the plane of
the binary.  The periapse of the forced eccentric orbit is aligned with
the periapse of the binary.  The free eccentricity and inclination
have the same meaning for a circumbinary orbit as in a Keplerian
system. There is precession of the free eccentric orbit in the
circumbinary case, for the same reason as in the case of a single,
oblate star (quadrupole and higher order contributions to
the time-averaged gravitational potential).

We place this analytical framework in the context of planet formation.
Our results demonstrate that (i) most circular orbits exist for
binaries even when their eccentricity is large enough that the theory
--- based on a perturbative approach --- becomes questionable; (ii)
non-intersecting, most circular orbits exist in the presence of a
massive disk and a large planet; (iii) these orbits remain apsidally
aligned with the binary --- if the binary precesses, the forced
epicyclic paths in the disk also precess; and (iv) the response of a
satellite to external perturbations (e.g., stirring from a distant
planet) is the same as it would be around a single star, in
the reference frame of a circular guiding center.  Thus, planets can
form {\it in situ} as close as a few times the orbital separation of
their binary hosts.  Nearer to home, satellites of the Pluto-Charon
binary may have formed in a similar way \citep{kb2014}.

\begin{figure}[htb]
\centerline{\includegraphics[width=7.0in]{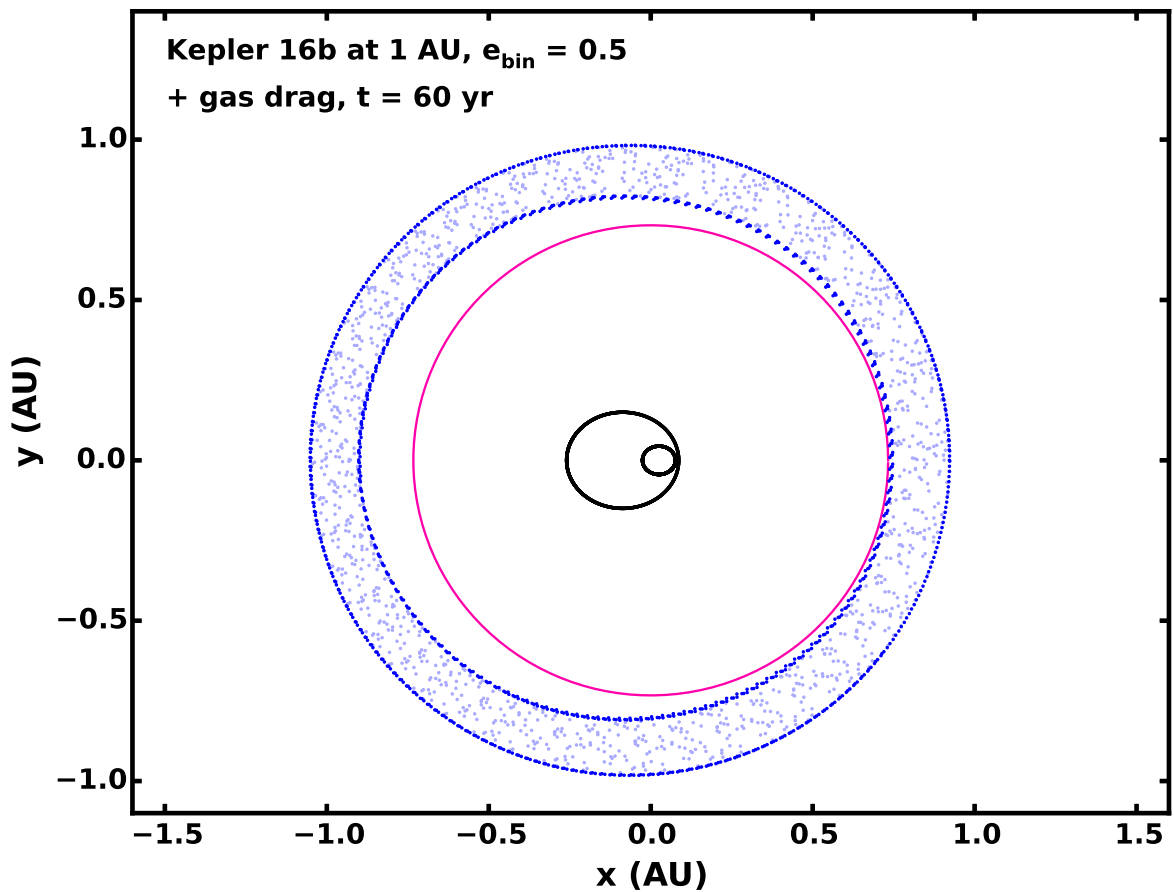}}
\caption{\label{fig:kepeccdragkep16ehix} Inspiral from gas drag. A
  sub-Keplerian gas disk has fluid elements on most circular orbits
  with $\eta = 0.001$ (Equation~(\ref{eq:eta})).  Embedded in it is a
  particle initially on a most circular orbit at 1~AU (outer blue
  ring). The particle evolves, drifting inward as a result of a drag
  force proportional to its speed relative to the gas (light blue
  points; the local gas speed is calculated using the Lee--Peale--Leung
   analytical theory). The particle's final orbit (inner blue ring) remains
  apsidally aligned with the binary, a pair of stars like Kepler-16
  except with $\ebin = 0.5$ (inner black curves). A circular path
  (magenta curve) provides a reference.} 
\end{figure}

\begin{figure}[htb]
\centerline{\includegraphics[width=7.0in]{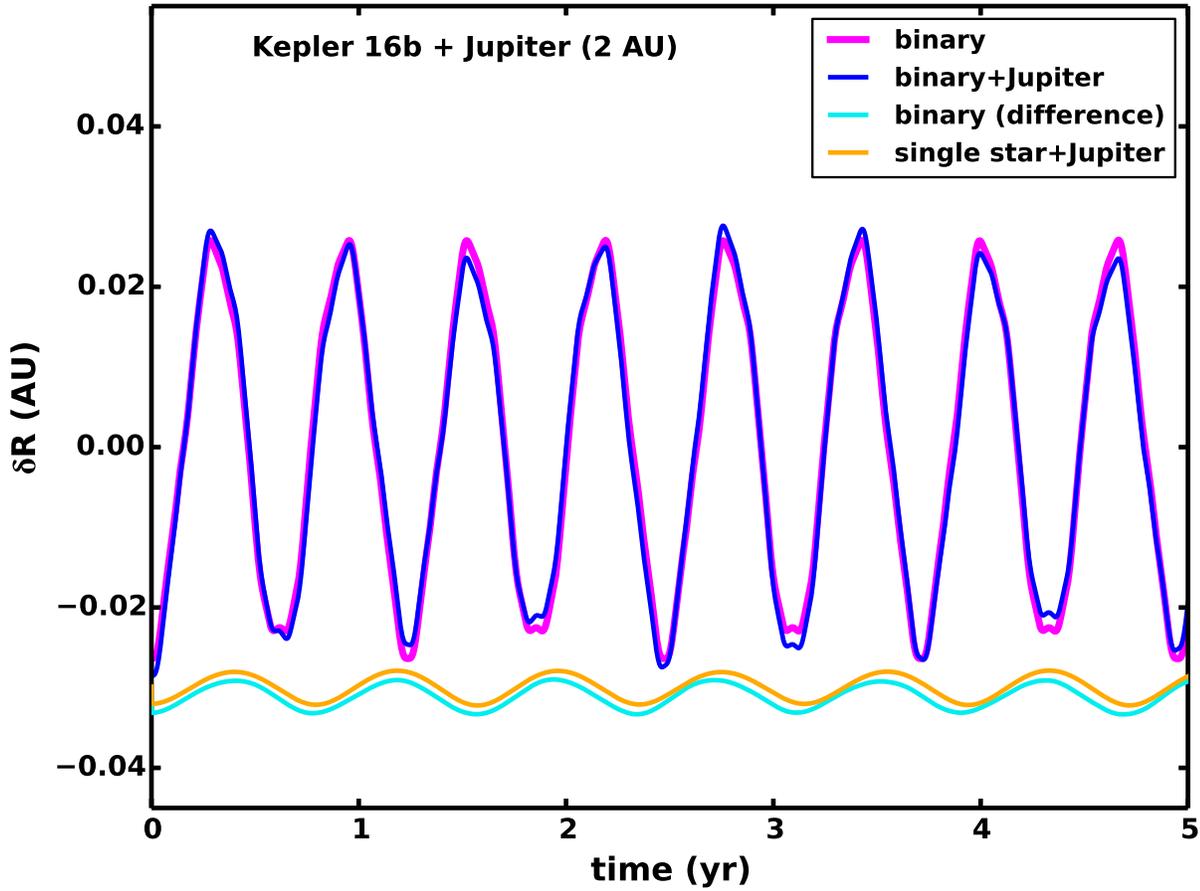}}
\caption{\label{fig:kepeccxkep16jupiter} The effect of an external perturber
  on circumbinary orbits. The curves show most circular paths around
  Kepler-16 at the orbital distance of Kepler-16b, both in isolation
  (magenta curve) and with a Jupiter-mass planet orbiting with
  $e\approx 0$ at 2~AU (blue curve). The difference between these two
  trajectories (cyan curve; offset for comparison) compares well with
  data from a satellite orbiting a single star with the mass of 
  Kepler-16 and a Jupiter-mass companion (orange curve). These
  lower curves show that the most circular path provides a frame of
  reference for the action of external perturbations, just like a
  circularly orbiting guiding center in the circumstellar case.}
\end{figure}

\begin{figure}[htb]
\centerline{\includegraphics[width=7.0in]{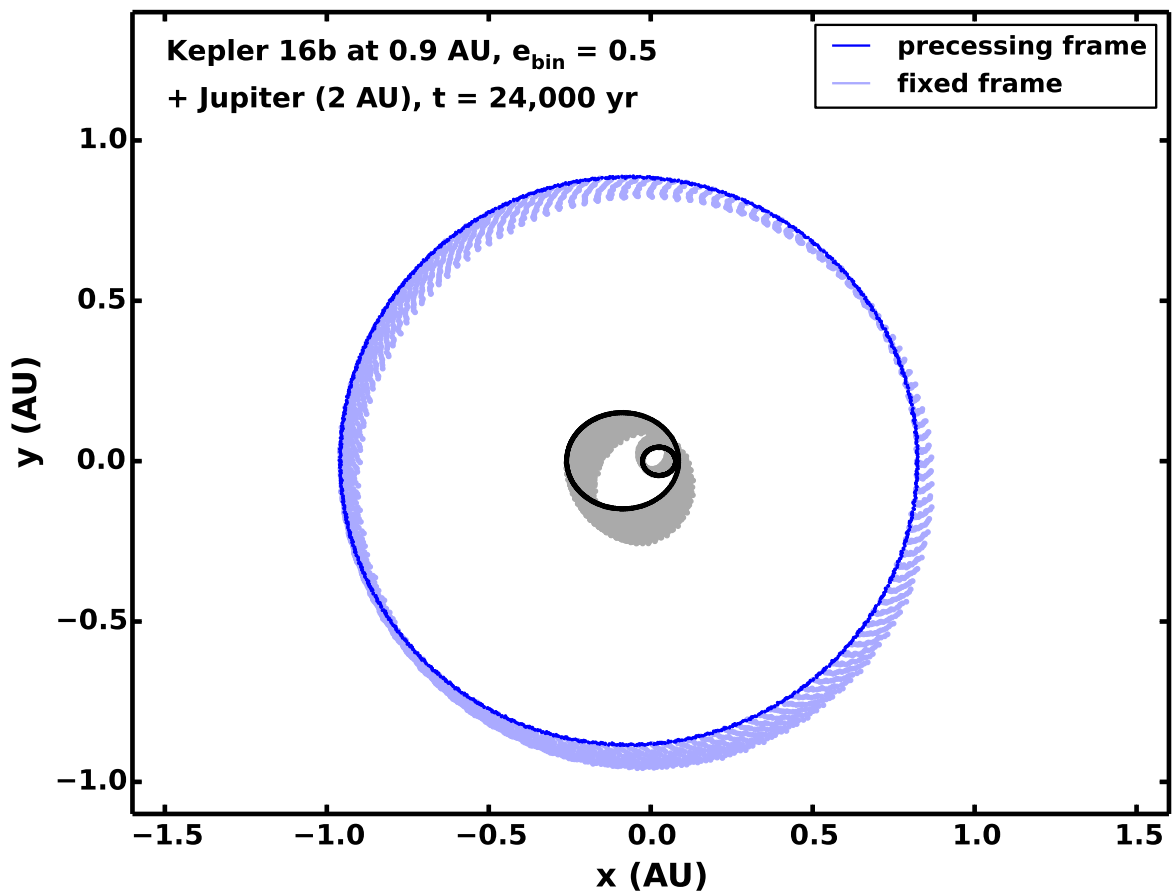}}
\caption{\label{fig:kepeccxkep16ehiwJx} Binary precession and
  circumbinary orbits in a simulation. The outer curves show most
  circular paths ($a = 0.9$~AU) around a binary like Kepler-16, set
  with eccentricity of 0.5 (inner curves) and with a Jupiter-mass
  planet orbiting with $e\approx 0$ at 2~AU. The gray and light blue-shaded curves show orbits in the inertial reference frame of the
  system's center of mass. The duration of the simulation is about a
  quarter of the precession period; orbits have precessed about
  90$^\circ$ (see Equation~(\ref{eq:precessjup})). The black and dark
  blue ellipses are the same orbits represented in a reference frame that
  precesses with the binary's periapse.  The satellite's forced
  epicyclic motion evidently precesses at this rate.}
\end{figure}

\begin{figure}[htb]
\centerline{\includegraphics[width=7.0in]{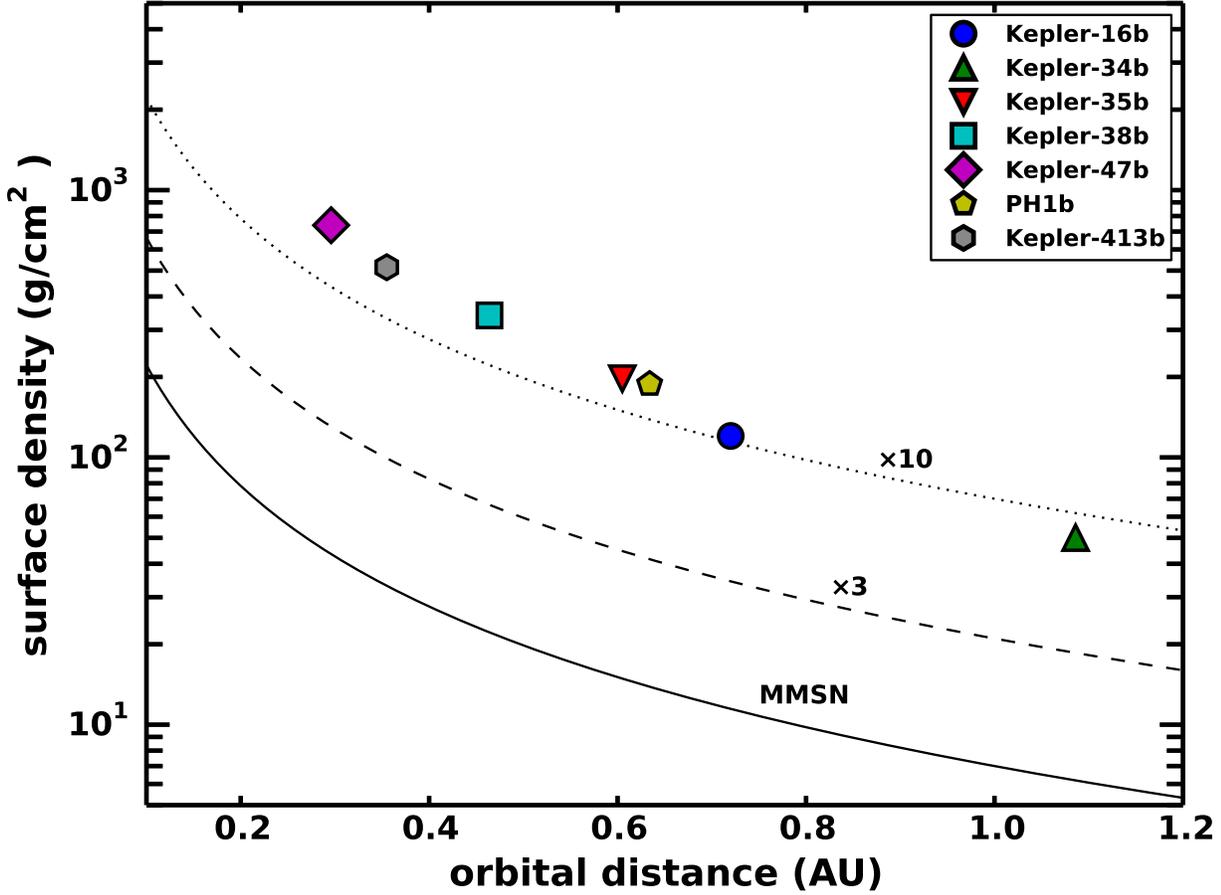}}
\caption{\label{fig:kepsigma} The minimum surface density to build the
  \kepler\ circumbinary planets.  Each planet is shown at its orbital
  distance from the host binary. The value of the surface density
  ($\Sigma$) comes from determining an annular width from which the
  planet could have accreted mass, based on its escape velocity when
  it had the mass of a 10~\Mearth, prior to its acquisition of a gas
  atmosphere (from Equation~(\ref{eq:feedzone})). The solid line is the
  surface density of a Minimum Mass Solar Nebula ($\Sigma = 7
  (a/\textrm{1\,AU})^{-1.5}$~g/cm$^2$), while the dashed and dotted
  lines correspond to disks with three and ten times that density (as
  labeled). Indications from simulations \citep[e.g.,][]{kb2006}
  suggest that the intermediate-mass disk is a realistic starting
  condition for the Solar nebula.}
\end{figure}

\begin{figure}[htb]
\centerline{\includegraphics[width=7.5in]{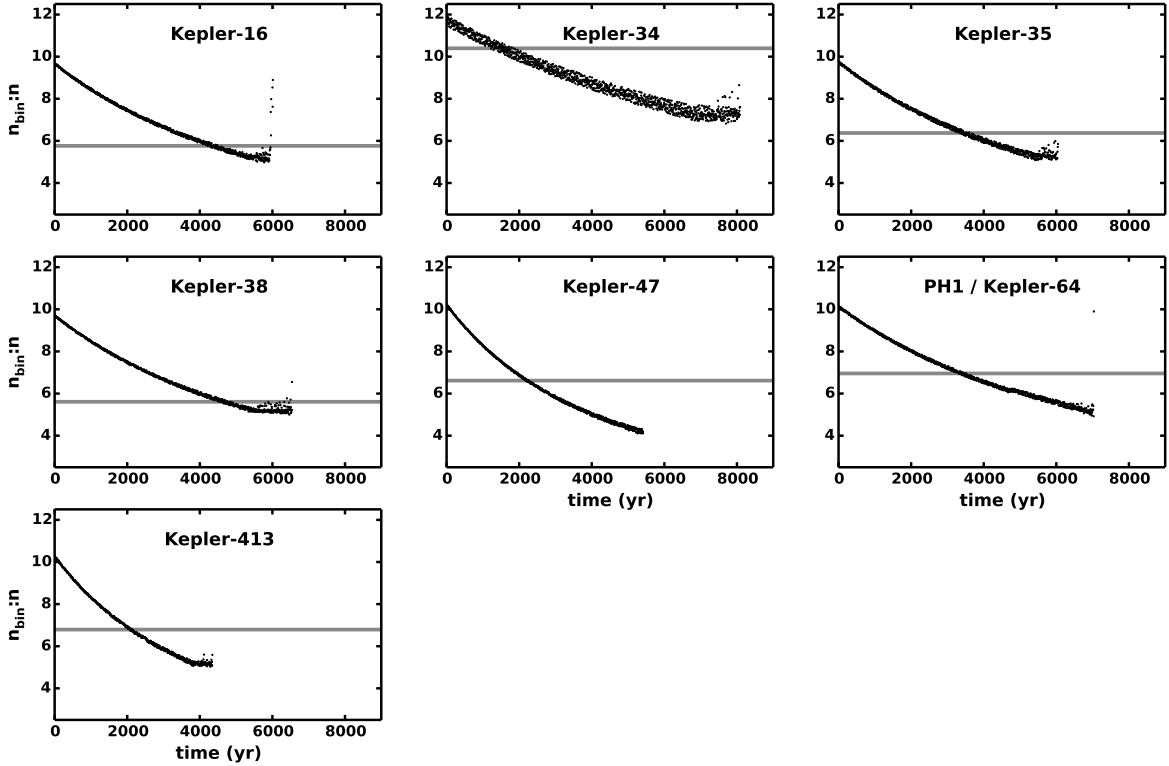}}
\caption{\label{fig:kepeccrez} Stability of circumbinary orbits.  Each
  panel shows the ratio of orbital frequencies between the central
  binary and the circumbinary planet.  To mimic migration, we smoothly
  adjust the binary semimajor axis to cover the observed ratio of
  orbital frequencies (gray line). The end of each curve indicates
  where the orbit of the planet becomes unstable. For most planets,
  the 5:1 resonance is disruptive.  Kepler-34b, whose binary host has
  high eccentricity ($e=0.5$), goes unstable when it hits the 7:1
  resonance.  The planet around Kepler-47, with the lowest binary
  eccentricity of the group ($e=0.023$), is stable down to the 4:1
  resonance.  Directly migrating the planet by artificially adjusting
  its semimajor axis gives similar results, except the planet around
  Kepler-34 remains bound.  Here, we choose to gradually expand the
  binary's semimajor axis while preserving all other orbital
  elements. It is mathematically equivalent, and it is more
  straightforward to adjust the Keplerian orbit of the binary than the
  non-Keplerian orbit of the planet.  }
\end{figure}

Extending these ideas to include gas within a protoplanetary disk, 
disk gravity and gas pressure slightly distort the shape of most 
circular orbits but not their alignment with the binary.
Although deviations from strictly circular geometry introduce new
hydrodynamical effects \citep[e.g.,][]{ogilvie2014, barker2014},
circumbinary gas streamlines plausibly settle on these most circular
paths as they do on circular orbits around a point mass. Interactions
with solid particles are also similar. The ``headwind'' felt by
planetesimals traveling through the gas is nearly the same as in a
circumstellar disk (Figure~\ref{fig:kepeccdragkep16ehix}). As the
gas dissipates in time, streamlines and planetesimals damp to the 
same set of most circular orbits. 

To explore whether our results are sensitive to instabilities of
circumbinary orbits, we also consider resonant excitations
\citep{arty1994,kley2014}. In our tests, the \kepler\ circumbinary
planets are all beyond the outermost unstable resonance around their
host. For most systems, this resonance is the 5:1 commensurability, 
close to the prediction of \citet{holman1999}. We define stability 
over a limited time frame ($10^3$--$10^5$~yr), not the age of the 
planetary systems. Nonetheless, our simulations suggest that with 
typical damping rates \citep{weiden1977b,gold2004,chiang2010} and
migration times \citep{ward1997}, both {\it in situ} formation and
migration of planets or their precursors are plausible.

Issues that we do not address include possible non-axisymmetric
structure in a massive circumbinary disk \citep{pel2013}, or the
physics near the edge of the stability zone at a distance of a few
times the binary separation. Simulations of gas and particle disks can
shed light on whether material is pushed outward by torque exchange
with the binary or lost in the unstable
zone \citep{arty1996,gunther2002}.  Furthermore, we do not consider
tidal evolution of the binary \citep[e.g.,][]{hurley2002}. Depending
on the nature of the evolution, variations in binary separation and
eccentricity will affect the stability of orbits in circumbinary
planetary system. For an intriguing example, see the discussion
by \citet{ward2006} on resonant transport of moons in orbits around the
Pluto-Charon binary as it tidally expands \citep{bk2015pc}.

Our results stand in contrast to previous theoretical studies of
circumbinary planet formation. The prevailing view is that dynamical
excitation of planetesimals by the binary leads to destructive
collisions
\citep{mori2004,meschiari2012,paarde2012,raf2013,xie2013,lines2014}.
In this interpretation, if the binary's eccentricity is even modestly
high, $\ebin \approx 0.2$, epicyclic velocities eventually exceed the
shattering speed for all but the largest planetesimals over a
significant range of orbital distances. 
Entrainment in a gas disk \citep[e.g.,][]{marzari2000}, or precession
from disk gravity \citep{raf2013, silsbee2015b} mitigate the situation
only for some range of planetesimal sizes or distant regions of a 
protoplanetary disk.  In this view, planet formation close to the 
binary is impossible.

These conclusions depend on the assumption that gas and planetesimals
are initially on orbits with no eccentricity in the Keplerian sense
\citep[see Figure~1 in][]{lines2014}. In light of analytical theory
(\S2), this choice endows particles with equal amounts of free and
forced eccentricity. As the free part drifts in phase, collisions
destroy growing protoplanets.  A more realistic approach to modeling
an unstirred disk is to set the gas and planetesimals on most circular
orbits with no free eccentricity.  Absent any stirring, planetesimals
in a particle disk orbit an eccentric binary host indefinitely
without colliding. If planetesimals develop small $e$ and $i$ about 
the most circular orbits, collision velocities are modest, just as for
planetesimals with modest $e$ and $i$ about a circular orbit around
a single star. Modest collision velocities promote growth instead of
destruction. Planet formation close to the binary is then robust.

Finally, we review characteristics of \kepler\ circumbinary planets in
light of our results. If circumbinary planets form roughly {\it in situ},
the orbital characteristics and sizes of these planets require very large
initial surface densities (10--20 times the MMSN). Although these high
surface densities appear to preclude {\it in situ} models with no migration,
the migrate-then-assemble picture of \citet{hansen2012} is viable. As
a plausible alternative, migrate-in-gas models allow these planets to form 
beyond the snow line and migrate inward through the gas. For this set of 
\kepler\ planets, migration through gas avoids unstable resonances around 
the binary. For Kepler-34b and Kepler-413b, formation beyond the snow line
followed by a scattering event is also a reasonable scenario.

In general, beyond the inner unstable cavity around a stellar binary,
the evolution of solids on most circular orbits differs little from
the evolution of solids on circular orbits around a single star. Thus,
the standard issues of formation around single stars (e.g., planetesimal
formation, migration,
resonances, scattering, etc) have clear parallels in circumbinary
disks \citep[for a summary of the issues in single stars,
  see][]{schlichting2014}.

We conclude with predictions for circumbinary planetary systems. In
our scenario, planets are as prevalent around binaries as around
single stars. Furthermore, relative to coplanar, most circular
orbits, these planets should have the same distribution of orbital
elements (free eccentricity and inclination) as their circumstellar
cousins.  Data from the full \kepler\ catalog indeed suggest that the
planets have comparable rates around binaries and single stars
\citep{armstrong2014}.  Tatooine sunsets may be common after all.

\acknowledgements

We thank 
the referee for a timely and thoughtful report that helped is to hone
our presentation. We also thank
N.\ Georgakarakos,
J.\ Pringle, 
R.\ Rafikov, 
P.\ Th\'{e}bault 
and
A.\ Youdin
for helpful
comments on the manuscript.  We gratefully acknowledge NASA for
support through the \textit{Astrophysics Theory} and \textit{Origins
  of Solar Systems} programs (grant NNX10AF35G) and through the
\textit{Outer Planets Program} (grant NNX11AM37G). We also acknowledge
NASA for a generous allocation of time on the 'discover'
supercomputing cluster.  This research has made use of the Exoplanet
Orbit Database and the Exoplanet Data Explorer at exoplanets.org.

\bibliography{planets}{}

\begin{thebibliography}{}
\expandafter\ifx\csname natexlab\endcsname\relax\def\natexlab#1{#1}\fi

\bibitem[{{Adachi} {et~al.}(1976){Adachi}, {Hayashi}, \& {Nakazawa}}]{ada76}
{Adachi}, I., {Hayashi}, C., \& {Nakazawa}, K. 1976, Progress of Theoretical
  Physics, 56, 1756

\bibitem[{{Andrews} {et~al.}(2013){Andrews}, {Rosenfeld}, {Kraus}, \&
  {Wilner}}]{and2013}
{Andrews}, S.~M., {Rosenfeld}, K.~A., {Kraus}, A.~L., \& {Wilner}, D.~J. 2013,
  \apj, 771, 129

\bibitem[{{Andrews} \& {Williams}(2005)}]{and2005}
{Andrews}, S.~M., \& {Williams}, J.~P. 2005, \apj, 631, 1134

\bibitem[{{Andrews} \& {Williams}(2007)}]{and2007}
---. 2007, \apj, 671, 1800

\bibitem[{{Armstrong} {et~al.}(2014){Armstrong}, {Osborn}, {Brown}, {Faedi},
  {G{\'o}mez Maqueo Chew}, {Martin}, {Pollacco}, \& {Udry}}]{armstrong2014}
{Armstrong}, D.~J., {Osborn}, H.~P., {Brown}, D.~J.~A., {et~al.} 2014, \mnras,
  444, 1873

\bibitem[{{Artymowicz} \& {Lubow}(1994)}]{arty1994}
{Artymowicz}, P., \& {Lubow}, S.~H. 1994, \apj, 421, 651

\bibitem[{{Artymowicz} \& {Lubow}(1996)}]{arty1996}
---. 1996, \apjl, 467, L77

\bibitem[{{Barker} \& {Ogilvie}(2014)}]{barker2014}
{Barker}, A.~J., \& {Ogilvie}, G.~I. 2014, \mnras, 445, 2637

\bibitem[{{Beaug{\'e}} \& {Nesvorn{\'y}}(2012)}]{beauge2012}
{Beaug{\'e}}, C., \& {Nesvorn{\'y}}, D. 2012, \apj, 751, 119

\bibitem[{{Benz} \& {Asphaug}(1999)}]{benz1999}
{Benz}, W., \& {Asphaug}, E. 1999, Icarus, 142, 5

\bibitem[{{Beuermann} {et~al.}(2011){Beuermann}, {Buhlmann}, {Diese},
  {Dreizler}, {Hessman}, {Husser}, {Miller}, {Nickol}, {Pons}, {Ruhr},
  {Schm{\"u}lling}, {Schwope}, {Sorge}, {Ulrichs}, {Winget}, \&
  {Winget}}]{Beuermann2011}
{Beuermann}, K., {Buhlmann}, J., {Diese}, J., {et~al.} 2011, \aap, 526, A53

\bibitem[{{Birnstiel} {et~al.}(2012){Birnstiel}, {Andrews}, \&
  {Ercolano}}]{birn2012}
{Birnstiel}, T., {Andrews}, S.~M., \& {Ercolano}, B. 2012, \aap, 544, A79

\bibitem[{{Birnstiel} {et~al.}(2010){Birnstiel}, {Dullemond}, \&
  {Brauer}}]{birn2010}
{Birnstiel}, T., {Dullemond}, C.~P., \& {Brauer}, F. 2010, \aap, 513, A79+

\bibitem[{{Borucki} {et~al.}(2011){Borucki}, {Koch}, {Basri}, {Batalha},
  {Brown}, {Bryson}, {Caldwell}, {Christensen-Dalsgaard}, {Cochran}, {DeVore},
  {Dunham}, {Gautier}, {Geary}, {Gilliland}, {Gould}, {Howell}, {Jenkins},
  {Latham}, {Lissauer}, {Marcy}, {Rowe}, {Sasselov}, {Boss}, {Charbonneau},
  {Ciardi}, {Doyle}, {Dupree}, {Ford}, {Fortney}, {Holman}, {Seager},
  {Steffen}, {Tarter}, {Welsh}, {Allen}, {Buchhave}, {Christiansen}, {Clarke},
  {Das}, {D{\'e}sert}, {Endl}, {Fabrycky}, {Fressin}, {Haas}, {Horch},
  {Howard}, {Isaacson}, {Kjeldsen}, {Kolodziejczak}, {Kulesa}, {Li}, {Lucas},
  {Machalek}, {McCarthy}, {MacQueen}, {Meibom}, {Miquel}, {Prsa}, {Quinn},
  {Quintana}, {Ragozzine}, {Sherry}, {Shporer}, {Tenenbaum}, {Torres},
  {Twicken}, {Van Cleve}, {Walkowicz}, {Witteborn}, \& {Still}}]{borucki2011}
{Borucki}, W.~J., {Koch}, D.~G., {Basri}, G., {et~al.} 2011, \apj, 736, 19

\bibitem[{{Brauer} {et~al.}(2008){Brauer}, {Dullemond}, \&
  {Henning}}]{brauer2008}
{Brauer}, F., {Dullemond}, C.~P., \& {Henning}, T. 2008, \aap, 480, 859

\bibitem[{{Bromley} \& {Kenyon}(2006)}]{bk2006}
{Bromley}, B.~C., \& {Kenyon}, S.~J. 2006, \aj, 131, 2737

\bibitem[{{Bromley} \& {Kenyon}(2011{\natexlab{a}})}]{bk2011a}
---. 2011{\natexlab{a}}, \apj, 731, 101

\bibitem[{{Bromley} \& {Kenyon}(2011{\natexlab{b}})}]{bk2011b}
---. 2011{\natexlab{b}}, \apj, 735, 29

\bibitem[{{Bromley} \& {Kenyon}(2014)}]{bk2014}
---. 2014, \apj, 796, 141

\bibitem[{{Bromley} \& {Kenyon}(2015)}]{bk2015pc}
---. 2015, ArXiv e-prints, arXiv:1503.06805

\bibitem[{{Burke} {et~al.}(2014){Burke}, {Bryson}, {Mullally}, {Rowe},
  {Christiansen}, {Thompson}, {Coughlin}, {Haas}, {Batalha}, {Caldwell},
  {Jenkins}, {Still}, {Barclay}, {Borucki}, {Chaplin}, {Ciardi}, {Clarke},
  {Cochran}, {Demory}, {Esquerdo}, {Gautier}, {Gilliland}, {Girouard}, {Havel},
  {Henze}, {Howell}, {Huber}, {Latham}, {Li}, {Morehead}, {Morton}, {Pepper},
  {Quintana}, {Ragozzine}, {Seader}, {Shah}, {Shporer}, {Tenenbaum}, {Twicken},
  \& {Wolfgang}}]{burke2014}
{Burke}, C.~J., {Bryson}, S.~T., {Mullally}, F., {et~al.} 2014, \apjs, 210, 19

\bibitem[{{Cassan} {et~al.}(2012){Cassan}, {Kubas}, {Beaulieu}, {Dominik},
  {Horne}, {Greenhill}, {Wambsganss}, {Menzies}, {Williams}, {J{\o}rgensen},
  {Udalski}, {Bennett}, {Albrow}, {Batista}, {Brillant}, {Caldwell}, {Cole},
  {Coutures}, {Cook}, {Dieters}, {Prester}, {Donatowicz}, {Fouqu{\'e}}, {Hill},
  {Kains}, {Kane}, {Marquette}, {Martin}, {Pollard}, {Sahu}, {Vinter},
  {Warren}, {Watson}, {Zub}, {Sumi}, {Szyma{\'n}ski}, {Kubiak}, {Poleski},
  {Soszynski}, {Ulaczyk}, {Pietrzy{\'n}ski}, \& {Wyrzykowski}}]{cassan2012}
{Cassan}, A., {Kubas}, D., {Beaulieu}, J.-P., {et~al.} 2012, \nat, 481, 167

\bibitem[{{Cassen} \& {Moosman}(1981)}]{cass1981}
{Cassen}, P., \& {Moosman}, A. 1981, Icarus, 48, 353

\bibitem[{{Chambers} \& {Wetherill}(1998)}]{chambers1998}
{Chambers}, J.~E., \& {Wetherill}, G.~W. 1998, Icarus, 136, 304

\bibitem[{{Chatterjee} {et~al.}(2008){Chatterjee}, {Ford}, {Matsumura}, \&
  {Rasio}}]{chatterjee2008}
{Chatterjee}, S., {Ford}, E.~B., {Matsumura}, S., \& {Rasio}, F.~A. 2008, \apj,
  686, 580

\bibitem[{{Chavez} {et~al.}(2015){Chavez}, {Georgakarakos}, {Prodan},
  {Reyes-Ruiz}, {Aceves}, {Betancourt}, \& {Perez-Tijerina}}]{chavez2015}
{Chavez}, C.~E., {Georgakarakos}, N., {Prodan}, S., {et~al.} 2015, \mnras, 446,
  1283

\bibitem[{{Cheng} {et~al.}(2014){Cheng}, {Peale}, \& {Lee}}]{cheng2014b}
{Cheng}, W.~H., {Peale}, S.~J., \& {Lee}, M.~H. 2014, \icarus, 241, 180

\bibitem[{{Chiang} \& {Youdin}(2010)}]{chiang2010}
{Chiang}, E., \& {Youdin}, A.~N. 2010, Annual Review of Earth and Planetary
  Sciences, 38, 493

\bibitem[{{Chiang} \& {Goldreich}(1997)}]{chiang1997}
{Chiang}, E.~I., \& {Goldreich}, P. 1997, \apj, 490, 368

\bibitem[{{Chirikov}(1959)}]{chirikov1959}
{Chirikov}, B.~V. 1959, Soviet Physics Doklady, 4, 390

\bibitem[{{Chirikov}(1979)}]{chirikov1979}
---. 1979, \physrep, 52, 263

\bibitem[{{Cumming} {et~al.}(2008){Cumming}, {Butler}, {Marcy}, {Vogt},
  {Wright}, \& {Fischer}}]{cum2008}
{Cumming}, A., {Butler}, R.~P., {Marcy}, G.~W., {et~al.} 2008, \pasp, 120, 531

\bibitem[{{Dauphas} \& {Chaussidon}(2011)}]{dauphas2011a}
{Dauphas}, N., \& {Chaussidon}, M. 2011, Annual Review of Earth and Planetary
  Sciences, 39, 351

\bibitem[{{Davis} {et~al.}(1985){Davis}, {Chapman}, {Weidenschilling}, \&
  {Greenberg}}]{davis1985}
{Davis}, D.~R., {Chapman}, C.~R., {Weidenschilling}, S.~J., \& {Greenberg}, R.
  1985, Icarus, 63, 30

\bibitem[{{Dent} {et~al.}(2013){Dent}, {Thi}, {Kamp}, {Williams}, {Menard},
  {Andrews}, {Ardila}, {Aresu}, {Augereau}, {Barrado y Navascues}, {Brittain},
  {Carmona}, {Ciardi}, {Danchi}, {Donaldson}, {Duchene}, {Eiroa}, {Fedele},
  {Grady}, {de Gregorio-Molsalvo}, {Howard}, {Hu{\'e}lamo}, {Krivov},
  {Lebreton}, {Liseau}, {Martin-Zaidi}, {Mathews}, {Meeus},
  {Mendigut{\'{\i}}a}, {Montesinos}, {Morales-Calderon}, {Mora}, {Nomura},
  {Pantin}, {Pascucci}, {Phillips}, {Pinte}, {Podio}, {Ramsay}, {Riaz},
  {Riviere-Marichalar}, {Roberge}, {Sandell}, {Solano}, {Tilling}, {Torrelles},
  {Vandenbusche}, {Vicente}, {White}, \& {Woitke}}]{dent2013}
{Dent}, W.~R.~F., {Thi}, W.~F., {Kamp}, I., {et~al.} 2013, \pasp, 125, 477

\bibitem[{{Dong} \& {Zhu}(2013)}]{dong2013}
{Dong}, S., \& {Zhu}, Z. 2013, \apj, 778, 53

\bibitem[{{Doolin} \& {Blundell}(2011)}]{doolin2011}
{Doolin}, S., \& {Blundell}, K.~M. 2011, \mnras, 418, 2656

\bibitem[{{Doyle} {et~al.}(2011){Doyle}, {Carter}, {Fabrycky}, {Slawson},
  {Howell}, {Winn}, {Orosz}, {Prsa}, {Welsh}, {Quinn}, {Latham}, {Torres},
  {Buchhave}, {Marcy}, {Fortney}, {Shporer}, {Ford}, {Lissauer}, {Ragozzine},
  {Rucker}, {Batalha}, {Jenkins}, {Borucki}, {Koch}, {Middour}, {Hall},
  {McCauliff}, {Fanelli}, {Quintana}, {Holman}, {Caldwell}, {Still},
  {Stefanik}, {Brown}, {Esquerdo}, {Tang}, {Furesz}, {Geary}, {Berlind},
  {Calkins}, {Short}, {Steffen}, {Sasselov}, {Dunham}, {Cochran}, {Boss},
  {Haas}, {Buzasi}, \& {Fischer}}]{doyle2011}
{Doyle}, L.~R., {Carter}, J.~A., {Fabrycky}, D.~C., {et~al.} 2011, Science,
  333, 1602

\bibitem[{{Garaud} {et~al.}(2013){Garaud}, {Meru}, {Galvagni}, \&
  {Olczak}}]{garaud2013}
{Garaud}, P., {Meru}, F., {Galvagni}, M., \& {Olczak}, C. 2013, \apj, 764, 146

\bibitem[{{Georgakarakos} \& {Eggl}(2015)}]{georgakarakos2015}
{Georgakarakos}, N., \& {Eggl}, S. 2015, \apj, 802, 94

\bibitem[{{Goldreich} {et~al.}(2004){Goldreich}, {Lithwick}, \&
  {Sari}}]{gold2004}
{Goldreich}, P., {Lithwick}, Y., \& {Sari}, R. 2004, \araa, 42, 549

\bibitem[{{Goldreich} \& {Tremaine}(1980)}]{gold1980}
{Goldreich}, P., \& {Tremaine}, S. 1980, \apj, 241, 425

\bibitem[{{Goldreich} \& {Tremaine}(1978)}]{gold1978}
{Goldreich}, P., \& {Tremaine}, S.~D. 1978, Icarus, 34, 227

\bibitem[{{Gould} {et~al.}(2010){Gould}, {Dong}, {Gaudi}, {Udalski}, {Bond},
  {Greenhill}, {Street}, {Dominik}, {Sumi}, {Szyma{\'n}ski}, {Han}, {Allen},
  {Bolt}, {Bos}, {Christie}, {DePoy}, {Drummond}, {Eastman}, {Gal-Yam},
  {Higgins}, {Janczak}, {Kaspi}, {Koz{\l}owski}, {Lee}, {Mallia}, {Maury},
  {Maoz}, {McCormick}, {Monard}, {Moorhouse}, {Morgan}, {Natusch}, {Ofek},
  {Park}, {Pogge}, {Polishook}, {Santallo}, {Shporer}, {Spector}, {Thornley},
  {Yee}, {{$\mu$}FUN Collaboration}, {Kubiak}, {Pietrzy{\'n}ski},
  {Soszy{\'n}ski}, {Szewczyk}, {Wyrzykowski}, {Ulaczyk}, {Poleski}, {OGLE
  Collaboration}, {Abe}, {Bennett}, {Botzler}, {Douchin}, {Freeman}, {Fukui},
  {Furusawa}, {Hearnshaw}, {Hosaka}, {Itow}, {Kamiya}, {Kilmartin}, {Korpela},
  {Lin}, {Ling}, {Makita}, {Masuda}, {Matsubara}, {Miyake}, {Muraki}, {Nagaya},
  {Nishimoto}, {Ohnishi}, {Okumura}, {Perrott}, {Philpott}, {Rattenbury},
  {Saito}, {Sako}, {Sullivan}, {Sweatman}, {Tristram}, {von Seggern}, {Yock},
  {MOA Collaboration}, {Albrow}, {Batista}, {Beaulieu}, {Brillant}, {Caldwell},
  {Calitz}, {Cassan}, {Cole}, {Cook}, {Coutures}, {Dieters}, {Dominis Prester},
  {Donatowicz}, {Fouqu{\'e}}, {Hill}, {Hoffman}, {Jablonski}, {Kane}, {Kains},
  {Kubas}, {Marquette}, {Martin}, {Martioli}, {Meintjes}, {Menzies},
  {Pedretti}, {Pollard}, {Sahu}, {Vinter}, {Wambsganss}, {Watson}, {Williams},
  {Zub}, {PLANET Collaboration}, {Allan}, {Bode}, {Bramich}, {Burgdorf},
  {Clay}, {Fraser}, {Hawkins}, {Horne}, {Kerins}, {Lister}, {Mottram},
  {Saunders}, {Snodgrass}, {Steele}, {Tsapras}, {RoboNet Collaboration},
  {J{\o}rgensen}, {Anguita}, {Bozza}, {Calchi Novati}, {Harps{\o}e}, {Hinse},
  {Hundertmark}, {Kj{\ae}rgaard}, {Liebig}, {Mancini}, {Masi}, {Mathiasen},
  {Rahvar}, {Ricci}, {Scarpetta}, {Southworth}, {Surdej}, {Th{\"o}ne}, \&
  {MiNDSTEp Consortium}}]{gould2010}
{Gould}, A., {Dong}, S., {Gaudi}, B.~S., {et~al.} 2010, \apj, 720, 1073

\bibitem[{{Greenberg} {et~al.}(1978){Greenberg}, {Hartmann}, {Chapman}, \&
  {Wacker}}]{green1978}
{Greenberg}, R., {Hartmann}, W.~K., {Chapman}, C.~R., \& {Wacker}, J.~F. 1978,
  Icarus, 35, 1

\bibitem[{{Greenzweig} \& {Lissauer}(1990)}]{green1990}
{Greenzweig}, Y., \& {Lissauer}, J.~J. 1990, Icarus, 87, 40

\bibitem[{{G{\"u}nther} \& {Kley}(2002)}]{gunther2002}
{G{\"u}nther}, R., \& {Kley}, W. 2002, \aap, 387, 550

\bibitem[{{Han} {et~al.}(2014){Han}, {Wang}, {Wright}, {Feng}, {Zhao},
  {Fakhouri}, {Brown}, \& {Hancock}}]{exoplanets2014}
{Han}, E., {Wang}, S.~X., {Wright}, J.~T., {et~al.} 2014, \pasp, 126, 827

\bibitem[{{Hansen}(2014)}]{hansen2014b}
{Hansen}, B.~M.~S. 2014, ArXiv e-prints, arXiv:1403.6553

\bibitem[{{Hansen} \& {Murray}(2012)}]{hansen2012}
{Hansen}, B.~M.~S., \& {Murray}, N. 2012, \apj, 751, 158

\bibitem[{{Harris} {et~al.}(2012){Harris}, {Andrews}, {Wilner}, \&
  {Kraus}}]{harris2012}
{Harris}, R.~J., {Andrews}, S.~M., {Wilner}, D.~J., \& {Kraus}, A.~L. 2012,
  \apj, 751, 115

\bibitem[{{Hayashi}(1981)}]{hayashi1981}
{Hayashi}, C. 1981, Progress of Theoretical Physics Supplement, 70, 35

\bibitem[{{Heppenheimer}(1978)}]{hepp1978}
{Heppenheimer}, T.~A. 1978, \aap, 65, 421

\bibitem[{{Holman} \& {Wiegert}(1999)}]{holman1999}
{Holman}, M.~J., \& {Wiegert}, P.~A. 1999, \aj, 117, 621

\bibitem[{{Hornung} {et~al.}(1985){Hornung}, {Pellat}, \& {Barge}}]{horn1985}
{Hornung}, P., {Pellat}, R., \& {Barge}, P. 1985, Icarus, 64, 295

\bibitem[{{Housen} \& {Holsapple}(1999)}]{housen1999}
{Housen}, K.~R., \& {Holsapple}, K.~A. 1999, Icarus, 142, 21

\bibitem[{{Howard} {et~al.}(2010){Howard}, {Marcy}, {Johnson}, {Fischer},
  {Wright}, {Isaacson}, {Valenti}, {Anderson}, {Lin}, \& {Ida}}]{how2010}
{Howard}, A.~W., {Marcy}, G.~W., {Johnson}, J.~A., {et~al.} 2010, Science, 330,
  653

\bibitem[{{Howard} {et~al.}(2012){Howard}, {Marcy}, {Bryson}, {Jenkins},
  {Rowe}, {Batalha}, {Borucki}, {Koch}, {Dunham}, {Gautier}, {Van Cleve},
  {Cochran}, {Latham}, {Lissauer}, {Torres}, {Brown}, {Gilliland}, {Buchhave},
  {Caldwell}, {Christensen-Dalsgaard}, {Ciardi}, {Fressin}, {Haas}, {Howell},
  {Kjeldsen}, {Seager}, {Rogers}, {Sasselov}, {Steffen}, {Basri},
  {Charbonneau}, {Christiansen}, {Clarke}, {Dupree}, {Fabrycky}, {Fischer},
  {Ford}, {Fortney}, {Tarter}, {Girouard}, {Holman}, {Johnson}, {Klaus},
  {Machalek}, {Moorhead}, {Morehead}, {Ragozzine}, {Tenenbaum}, {Twicken},
  {Quinn}, {Isaacson}, {Shporer}, {Lucas}, {Walkowicz}, {Welsh}, {Boss},
  {Devore}, {Gould}, {Smith}, {Morris}, {Prsa}, {Morton}, {Still}, {Thompson},
  {Mullally}, {Endl}, \& {MacQueen}}]{how2012}
{Howard}, A.~W., {Marcy}, G.~W., {Bryson}, S.~T., {et~al.} 2012, \apjs, 201, 15

\bibitem[{{Hurley} {et~al.}(2002){Hurley}, {Tout}, \& {Pols}}]{hurley2002}
{Hurley}, J.~R., {Tout}, C.~A., \& {Pols}, O.~R. 2002, \mnras, 329, 897

\bibitem[{{Juri{\'c}} \& {Tremaine}(2008)}]{juric2008}
{Juri{\'c}}, M., \& {Tremaine}, S. 2008, \apj, 686, 603

\bibitem[{{Kennedy}(2015)}]{kennedy2015}
{Kennedy}, G.~M. 2015, \mnras, 447, L75

\bibitem[{{Kennedy} \& {Kenyon}(2008)}]{kk2008}
{Kennedy}, G.~M., \& {Kenyon}, S.~J. 2008, \apj, 673, 502

\bibitem[{{Kennedy} {et~al.}(2012){Kennedy}, {Wyatt}, {Sibthorpe}, {Phillips},
  {Matthews}, \& {Greaves}}]{kennedy2012}
{Kennedy}, G.~M., {Wyatt}, M.~C., {Sibthorpe}, B., {et~al.} 2012, \mnras, 426,
  2115

\bibitem[{{Kenyon} \& {Bromley}(2001)}]{kb2001}
{Kenyon}, S.~J., \& {Bromley}, B.~C. 2001, \aj, 121, 538

\bibitem[{{Kenyon} \& {Bromley}(2002)}]{kb2002a}
---. 2002, \aj, 123, 1757

\bibitem[{{Kenyon} \& {Bromley}(2004)}]{kb2004b}
---. 2004, \apjl, 602, L133

\bibitem[{{Kenyon} \& {Bromley}(2005)}]{kb2005}
---. 2005, \aj, 130, 269

\bibitem[{{Kenyon} \& {Bromley}(2006)}]{kb2006}
---. 2006, \aj, 131, 1837

\bibitem[{{Kenyon} \& {Bromley}(2008)}]{kb2008}
---. 2008, \apjs, 179, 451

\bibitem[{{Kenyon} \& {Bromley}(2010)}]{kb2010}
---. 2010, \apjs, 188, 242

\bibitem[{{Kenyon} \& {Bromley}(2014)}]{kb2014}
---. 2014, \aj, 147, 8

\bibitem[{{Kenyon} \& {Bromley}(2015)}]{kb2015}
---. 2015, ArXiv e-prints, arXiv:1501.05659

\bibitem[{{Kenyon} \& {Luu}(1998)}]{kl1998}
{Kenyon}, S.~J., \& {Luu}, J.~X. 1998, \aj, 115, 2136

\bibitem[{{Kleine} {et~al.}(2009){Kleine}, {Touboul}, {Bourdon}, {Nimmo},
  {Mezger}, {Palme}, {Jacobsen}, {Yin}, \& {Halliday}}]{kleine2009}
{Kleine}, T., {Touboul}, M., {Bourdon}, B., {et~al.} 2009, \gca, 73, 5150

\bibitem[{{Kley} \& {Haghighipour}(2014)}]{kley2014}
{Kley}, W., \& {Haghighipour}, N. 2014, \aap, 564, A72

\bibitem[{{Kostov} {et~al.}(2013){Kostov}, {McCullough}, {Hinse}, {Tsvetanov},
  {H{\'e}brard}, {D{\'{\i}}az}, {Deleuil}, \& {Valenti}}]{kostov2013}
{Kostov}, V.~B., {McCullough}, P.~R., {Hinse}, T.~C., {et~al.} 2013, \apj, 770,
  52

\bibitem[{{Kostov} {et~al.}(2014){Kostov}, {McCullough}, {Carter}, {Deleuil},
  {D{\'{\i}}az}, {Fabrycky}, {H{\'e}brard}, {Hinse}, {Mazeh}, {Orosz},
  {Tsvetanov}, \& {Welsh}}]{kostov2014}
{Kostov}, V.~B., {McCullough}, P.~R., {Carter}, J.~A., {et~al.} 2014, \apj,
  784, 14

\bibitem[{{Laibe} {et~al.}(2012){Laibe}, {Gonzalez}, \& {Maddison}}]{laibe2012}
{Laibe}, G., {Gonzalez}, J.-F., \& {Maddison}, S.~T. 2012, \aap, 537, A61

\bibitem[{{Lecar} {et~al.}(2001){Lecar}, {Franklin}, {Holman}, \&
  {Murray}}]{lecar2001}
{Lecar}, M., {Franklin}, F.~A., {Holman}, M.~J., \& {Murray}, N.~J. 2001,
  \araa, 39, 581

\bibitem[{{Lee} \& {Peale}(2006)}]{lee2006}
{Lee}, M.~H., \& {Peale}, S.~J. 2006, \icarus, 184, 573

\bibitem[{{Leung} \& {Lee}(2013)}]{leung2013}
{Leung}, G.~C.~K., \& {Lee}, M.~H. 2013, \apj, 763, 107

\bibitem[{{Lightman}(1974)}]{lightman1974}
{Lightman}, A.~P. 1974, \apj, 194, 419

\bibitem[{{Lin} \& {Papaloizou}(1979)}]{lin1979b}
{Lin}, D.~N.~C., \& {Papaloizou}, J. 1979, \mnras, 188, 191

\bibitem[{{Lin} \& {Pringle}(1976)}]{lin1976}
{Lin}, D.~N.~C., \& {Pringle}, J.~E. 1976, in IAU Symposium, Vol.~73, Structure
  and Evolution of Close Binary Systems, ed. P.~{Eggleton}, S.~{Mitton}, \&
  J.~{Whelan}, 237

\bibitem[{{Lines} {et~al.}(2014){Lines}, {Leinhardt}, {Paardekooper},
  {Baruteau}, \& {Thebault}}]{lines2014}
{Lines}, S., {Leinhardt}, Z.~M., {Paardekooper}, S., {Baruteau}, C., \&
  {Thebault}, P. 2014, \apjl, 782, L11

\bibitem[{{Lissauer}(1987)}]{liss1987}
{Lissauer}, J.~J. 1987, Icarus, 69, 249

\bibitem[{{Lithwick} \& {Wu}(2008)}]{lith2008a}
{Lithwick}, Y., \& {Wu}, Y. 2008, ArXiv e-prints, arXiv:0802.2939

\bibitem[{{Lynden-Bell} \& {Pringle}(1974)}]{lbp1974}
{Lynden-Bell}, D., \& {Pringle}, J.~E. 1974, \mnras, 168, 603

\bibitem[{{Macintosh} {et~al.}(2014){Macintosh}, {Graham}, {Ingraham},
  {Konopacky}, {Marois}, {Perrin}, {Poyneer}, {Bauman}, {Barman}, {Burrows},
  {Cardwell}, {Chilcote}, {De Rosa}, {Dillon}, {Doyon}, {Dunn}, {Erikson},
  {Fitzgerald}, {Gavel}, {Goodsell}, {Hartung}, {Hibon}, {Kalas}, {Larkin},
  {Maire}, {Marchis}, {Marley}, {McBride}, {Millar-Blanchaer}, {Morzinski},
  {Norton}, {Oppenheimer}, {Palmer}, {Patience}, {Pueyo}, {Rantakyro},
  {Sadakuni}, {Saddlemyer}, {Savransky}, {Serio}, {Soummer},
  {Sivaramakrishnan}, {Song}, {Thomas}, {Wallace}, {Wiktorowicz}, \&
  {Wolff}}]{mac2014}
{Macintosh}, B., {Graham}, J.~R., {Ingraham}, P., {et~al.} 2014, Proceedings of
  the National Academy of Science, 111, 12661

\bibitem[{{Marzari} {et~al.}(2010){Marzari}, {Baruteau}, \&
  {Scholl}}]{marzari2010}
{Marzari}, F., {Baruteau}, C., \& {Scholl}, H. 2010, \aap, 514, L4

\bibitem[{{Marzari} \& {Scholl}(2000)}]{marzari2000}
{Marzari}, F., \& {Scholl}, H. 2000, \apj, 543, 328

\bibitem[{{Mayor} {et~al.}(2014){Mayor}, {Lovis}, \& {Santos}}]{mayor2014}
{Mayor}, M., {Lovis}, C., \& {Santos}, N.~C. 2014, \nat, 513, 328

\bibitem[{{Mayor} {et~al.}(2011){Mayor}, {Marmier}, {Lovis}, {Udry},
  {S{\'e}gransan}, {Pepe}, {Benz}, {Bertaux}, {Bouchy}, {Dumusque}, {Lo Curto},
  {Mordasini}, {Queloz}, \& {Santos}}]{mayor2011}
{Mayor}, M., {Marmier}, M., {Lovis}, C., {et~al.} 2011, ArXiv e-prints,
  arXiv:1109.2497

\bibitem[{{Meibom} \& {Mathieu}(2005)}]{meibom2005}
{Meibom}, S., \& {Mathieu}, R.~D. 2005, \apj, 620, 970

\bibitem[{{Meschiari}(2012)}]{meschiari2012}
{Meschiari}, S. 2012, \apj, 752, 71

\bibitem[{{Meschiari}(2014)}]{meschiari2014}
---. 2014, \apj, 790, 41

\bibitem[{{Moeckel} \& {Armitage}(2012)}]{moeckel2012}
{Moeckel}, N., \& {Armitage}, P.~J. 2012, \mnras, 419, 366

\bibitem[{{Moriwaki} \& {Nakagawa}(2004)}]{mori2004}
{Moriwaki}, K., \& {Nakagawa}, Y. 2004, \apj, 609, 1065

\bibitem[{{Mullally} {et~al.}(2015){Mullally}, {Coughlin}, {Thompson}, {Rowe},
  {Burke}, {Latham}, {Batalha}, {Bryson}, {Christiansen}, {Henze}, {Ofir},
  {Quarles}, {Shporer}, {Van Eylen}, {Van Laerhoven}, {Shah}, {Wolfgang},
  {Chaplin}, {Xie}, {Akeson}, {Argabright}, {Bachtell}, {Barclay}, {Borucki},
  {Caldwell}, {Campbell}, {Catanzarite}, {Cochran}, {Duren}, {Fleming},
  {Fraquelli}, {Girouard}, {Haas}, {He{\l}miniak}, {Howell}, {Huber}, {Larson},
  {Gautier}, {Jenkins}, {Li}, {Lissauer}, {McArthur}, {Miller}, {Morris},
  {Patil-Sabale}, {Plavchan}, {Putnam}, {Quintana}, {Ramirez}, {Silva Aguirre},
  {Seader}, {Smith}, {Steffen}, {Stewart}, {Stober}, {Still}, {Tenenbaum},
  {Troeltzsch}, {Twicken}, \& {Zamudio}}]{mullally2015}
{Mullally}, F., {Coughlin}, J.~L., {Thompson}, S.~E., {et~al.} 2015, \apjs,
  217, 31

\bibitem[{{Murray} \& {Dermott}(1999)}]{murray1999}
{Murray}, C.~D., \& {Dermott}, S.~F. 1999, {Solar system dynamics} (Princeton:
  Princeton University Press)

\bibitem[{{Musielak} {et~al.}(2005){Musielak}, {Cuntz}, {Marshall}, \&
  {Stuit}}]{musielak2005}
{Musielak}, Z.~E., {Cuntz}, M., {Marshall}, E.~A., \& {Stuit}, T.~D. 2005,
  \aap, 434, 355

\bibitem[{{Najita} \& {Kenyon}(2014)}]{najita2014}
{Najita}, J.~R., \& {Kenyon}, S.~J. 2014, \mnras, 445, 3315

\bibitem[{{Ogilvie}(2001)}]{ogilvie2001}
{Ogilvie}, G.~I. 2001, \mnras, 325, 231

\bibitem[{{Ogilvie} \& {Barker}(2014)}]{ogilvie2014}
{Ogilvie}, G.~I., \& {Barker}, A.~J. 2014, \mnras, 445, 2621

\bibitem[{{Ohtsuki}(1999)}]{oht1999}
{Ohtsuki}, K. 1999, \icarus, 137, 152

\bibitem[{{Ohtsuki} {et~al.}(2002){Ohtsuki}, {Stewart}, \& {Ida}}]{oht2002}
{Ohtsuki}, K., {Stewart}, G.~R., \& {Ida}, S. 2002, Icarus, 155, 436

\bibitem[{{Orosz} {et~al.}(2012{\natexlab{a}}){Orosz}, {Welsh}, {Carter},
  {Fabrycky}, {Cochran}, {Endl}, {Ford}, {Haghighipour}, {MacQueen}, {Mazeh},
  {Sanchis-Ojeda}, {Short}, {Torres}, {Agol}, {Buchhave}, {Doyle}, {Isaacson},
  {Lissauer}, {Marcy}, {Shporer}, {Windmiller}, {Barclay}, {Boss}, {Clarke},
  {Fortney}, {Geary}, {Holman}, {Huber}, {Jenkins}, {Kinemuchi}, {Kruse},
  {Ragozzine}, {Sasselov}, {Still}, {Tenenbaum}, {Uddin}, {Winn}, {Koch}, \&
  {Borucki}}]{orosz2012a}
{Orosz}, J.~A., {Welsh}, W.~F., {Carter}, J.~A., {et~al.} 2012{\natexlab{a}},
  Science, 337, 1511

\bibitem[{{Orosz} {et~al.}(2012{\natexlab{b}}){Orosz}, {Welsh}, {Carter},
  {Brugamyer}, {Buchhave}, {Cochran}, {Endl}, {Ford}, {MacQueen}, {Short},
  {Torres}, {Windmiller}, {Agol}, {Barclay}, {Caldwell}, {Clarke}, {Doyle},
  {Fabrycky}, {Geary}, {Haghighipour}, {Holman}, {Ibrahim}, {Jenkins},
  {Kinemuchi}, {Li}, {Lissauer}, {Pr{\v s}a}, {Ragozzine}, {Shporer}, {Still},
  \& {Wade}}]{orosz2012b}
---. 2012{\natexlab{b}}, \apj, 758, 87

\bibitem[{{Paardekooper} {et~al.}(2012){Paardekooper}, {Leinhardt},
  {Th{\'e}bault}, \& {Baruteau}}]{paarde2012}
{Paardekooper}, S.-J., {Leinhardt}, Z.~M., {Th{\'e}bault}, P., \& {Baruteau},
  C. 2012, \apjl, 754, L16

\bibitem[{{Pelupessy} \& {Portegies Zwart}(2013)}]{pel2013}
{Pelupessy}, F.~I., \& {Portegies Zwart}, S. 2013, \mnras, 429, 895

\bibitem[{{Pichardo} {et~al.}(2005){Pichardo}, {Sparke}, \&
  {Aguilar}}]{pichardo2005}
{Pichardo}, B., {Sparke}, L.~S., \& {Aguilar}, L.~A. 2005, \mnras, 359, 521

\bibitem[{{Pichardo} {et~al.}(2008){Pichardo}, {Sparke}, \&
  {Aguilar}}]{pichardo2008}
---. 2008, \mnras, 391, 815

\bibitem[{{Pierens} \& {Nelson}(2007)}]{pierens2007}
{Pierens}, A., \& {Nelson}, R.~P. 2007, \aap, 472, 993

\bibitem[{{Pierens} \& {Nelson}(2008{\natexlab{a}})}]{pierens2008a}
---. 2008{\natexlab{a}}, \aap, 478, 939

\bibitem[{{Pierens} \& {Nelson}(2008{\natexlab{b}})}]{pierens2008b}
---. 2008{\natexlab{b}}, \aap, 483, 633

\bibitem[{{Pinte} \& {Laibe}(2014)}]{pinte2014}
{Pinte}, C., \& {Laibe}, G. 2014, \aap, 565, A129

\bibitem[{{Piso} {et~al.}(2015){Piso}, {Youdin}, \& {Murray-Clay}}]{piso2015}
{Piso}, A.-M.~A., {Youdin}, A.~N., \& {Murray-Clay}, R.~A. 2015, \apj, 800, 82

\bibitem[{{Pollack} {et~al.}(1996){Pollack}, {Hubickyj}, {Bodenheimer},
  {Lissauer}, {Podolak}, \& {Greenzweig}}]{pollack1996}
{Pollack}, J.~B., {Hubickyj}, O., {Bodenheimer}, P., {et~al.} 1996, Icarus,
  124, 62

\bibitem[{{Popova} \& {Shevchenko}(2013)}]{popova2013}
{Popova}, E.~A., \& {Shevchenko}, I.~I. 2013, \apj, 769, 152

\bibitem[{{Pringle}(1981)}]{pri1981}
{Pringle}, J.~E. 1981, \araa, 19, 137

\bibitem[{{Pringle}(1991)}]{pri1991}
---. 1991, \mnras, 248, 754

\bibitem[{{Quintana} \& {Lissauer}(2006)}]{quintana2006}
{Quintana}, E.~V., \& {Lissauer}, J.~J. 2006, \icarus, 185, 1

\bibitem[{{Rafikov}(2004)}]{raf2004}
{Rafikov}, R.~R. 2004, \aj, 128, 1348

\bibitem[{{Rafikov}(2011)}]{raf2011}
---. 2011, \apj, 727, 86

\bibitem[{{Rafikov}(2013)}]{raf2013}
---. 2013, \apjl, 764, L16

\bibitem[{{Rasio} \& {Ford}(1996)}]{rasio1996}
{Rasio}, F.~A., \& {Ford}, E.~B. 1996, Science, 274, 954

\bibitem[{{Raymond} {et~al.}(2010){Raymond}, {Armitage}, \&
  {Gorelick}}]{raymond2010}
{Raymond}, S.~N., {Armitage}, P.~J., \& {Gorelick}, N. 2010, \apj, 711, 772

\bibitem[{{Raymond} {et~al.}(2006){Raymond}, {Quinn}, \&
  {Lunine}}]{raymond2006}
{Raymond}, S.~N., {Quinn}, T., \& {Lunine}, J.~I. 2006, \icarus, 183, 265

\bibitem[{{Raymond} {et~al.}(2011){Raymond}, {Armitage}, {Moro-Mart{\'{\i}}n},
  {Booth}, {Wyatt}, {Armstrong}, {Mandell}, {Selsis}, \& {West}}]{raymond2011}
{Raymond}, S.~N., {Armitage}, P.~J., {Moro-Mart{\'{\i}}n}, A., {et~al.} 2011,
  \aap, 530, A62

\bibitem[{{Rogers} {et~al.}(2011){Rogers}, {Bodenheimer}, {Lissauer}, \&
  {Seager}}]{rogers2011}
{Rogers}, L.~A., {Bodenheimer}, P., {Lissauer}, J.~J., \& {Seager}, S. 2011,
  \apj, 738, 59

\bibitem[{{Safronov}(1969)}]{saf1969}
{Safronov}, V.~S. 1969, {Evoliutsiia doplanetnogo oblaka. (Evolution of the
  Protoplanetary Cloud and Formation of the Earth and Planets, Nauka, Moscow
  [Translation 1972, NASA TT F-677]} (1969.)

\bibitem[{{Schlichting}(2014)}]{schlichting2014}
{Schlichting}, H.~E. 2014, \apjl, 795, L15

\bibitem[{{Schneider} {et~al.}(2011){Schneider}, {Dedieu}, {Le Sidaner},
  {Savalle}, \& {Zolotukhin}}]{schneider2011}
{Schneider}, J., {Dedieu}, C., {Le Sidaner}, P., {Savalle}, R., \&
  {Zolotukhin}, I. 2011, \aap, 532, A79

\bibitem[{{Scholl} {et~al.}(2007){Scholl}, {Marzari}, \&
  {Th{\'e}bault}}]{scholl2007}
{Scholl}, H., {Marzari}, F., \& {Th{\'e}bault}, P. 2007, \mnras, 380, 1119

\bibitem[{{Schwamb} {et~al.}(2013){Schwamb}, {Orosz}, {Carter}, {Welsh},
  {Fischer}, {Torres}, {Howard}, {Crepp}, {Keel}, {Lintott}, {Kaib}, {Terrell},
  {Gagliano}, {Jek}, {Parrish}, {Smith}, {Lynn}, {Simpson}, {Giguere}, \&
  {Schawinski}}]{schwamb2013}
{Schwamb}, M.~E., {Orosz}, J.~A., {Carter}, J.~A., {et~al.} 2013, \apj, 768,
  127

\bibitem[{{Silsbee} \& {Rafikov}(2015{\natexlab{a}})}]{silsbee2015b}
{Silsbee}, K., \& {Rafikov}, R.~R. 2015{\natexlab{a}}, ArXiv e-prints,
  arXiv:1504.00460

\bibitem[{{Silsbee} \& {Rafikov}(2015{\natexlab{b}})}]{silsbee2015a}
---. 2015{\natexlab{b}}, \apj, 798, 71

\bibitem[{{Spaute} {et~al.}(1991){Spaute}, {Weidenschilling}, {Davis}, \&
  {Marzari}}]{spaute1991}
{Spaute}, D., {Weidenschilling}, S.~J., {Davis}, D.~R., \& {Marzari}, F. 1991,
  Icarus, 92, 147

\bibitem[{{Stewart} \& {Wetherill}(1988)}]{stewart1988}
{Stewart}, G.~R., \& {Wetherill}, G.~W. 1988, \icarus, 74, 542

\bibitem[{{Syer} \& {Clarke}(1992)}]{syer1992}
{Syer}, D., \& {Clarke}, C.~J. 1992, \mnras, 255, 92

\bibitem[{{Szebehely}(1967)}]{szeb1967}
{Szebehely}, V. 1967, {Theory of orbits. The restricted problem of three
  bodies} (Academic Press, New York, NY)

\bibitem[{{Tamura}(2014)}]{tamura2014}
{Tamura}, M. 2014, in American Astronomical Society Meeting Abstracts, Vol.
  224, American Astronomical Society Meeting Abstracts 224, 301.03

\bibitem[{{Tanaka} {et~al.}(2002){Tanaka}, {Takeuchi}, \& {Ward}}]{tanaka2002}
{Tanaka}, H., {Takeuchi}, T., \& {Ward}, W.~R. 2002, \apj, 565, 1257

\bibitem[{{Terebey} {et~al.}(1984){Terebey}, {Shu}, \& {Cassen}}]{tsc1984}
{Terebey}, S., {Shu}, F.~H., \& {Cassen}, P. 1984, \apj, 286, 529

\bibitem[{{Th{\'e}bault} {et~al.}(2006){Th{\'e}bault}, {Marzari}, \&
  {Scholl}}]{thebault2006}
{Th{\'e}bault}, P., {Marzari}, F., \& {Scholl}, H. 2006, \icarus, 183, 193

\bibitem[{{Trilling} {et~al.}(2007){Trilling}, {Stansberry}, {Stapelfeldt},
  {Rieke}, {Su}, {Gray}, {Corbally}, {Bryden}, {Chen}, {Boden}, \&
  {Beichman}}]{trill2007}
{Trilling}, D.~E., {Stansberry}, J.~A., {Stapelfeldt}, K.~R., {et~al.} 2007,
  \apj, 658, 1289

\bibitem[{{Ward}(1981)}]{ward1981}
{Ward}, W.~R. 1981, Icarus, 47, 234

\bibitem[{{Ward}(1997)}]{ward1997}
---. 1997, \icarus, 126, 261

\bibitem[{{Ward} \& {Canup}(2006)}]{ward2006}
{Ward}, W.~R., \& {Canup}, R.~M. 2006, Science, 313, 1107

\bibitem[{{Weidenschilling}(1977{\natexlab{a}})}]{weiden1977a}
{Weidenschilling}, S.~J. 1977{\natexlab{a}}, \mnras, 180, 57

\bibitem[{{Weidenschilling}(1977{\natexlab{b}})}]{weiden1977b}
---. 1977{\natexlab{b}}, \apss, 51, 153

\bibitem[{{Weidenschilling}(1989)}]{weiden1989}
---. 1989, Icarus, 80, 179

\bibitem[{{Weidenschilling} \& {Cuzzi}(1993)}]{weiden1993}
{Weidenschilling}, S.~J., \& {Cuzzi}, J.~N. 1993, in Protostars and Planets
  III, ed. E.~H. {Levy} \& J.~I. {Lunine} (University of Arizona Press, Tucson,
  AZ), 1031--1060

\bibitem[{{Welsh} {et~al.}(2012){Welsh}, {Orosz}, {Carter}, {Fabrycky}, {Ford},
  {Lissauer}, {Pr{\v s}a}, {Quinn}, {Ragozzine}, {Short}, {Torres}, {Winn},
  {Doyle}, {Barclay}, {Batalha}, {Bloemen}, {Brugamyer}, {Buchhave},
  {Caldwell}, {Caldwell}, {Christiansen}, {Ciardi}, {Cochran}, {Endl},
  {Fortney}, {Gautier}, {Gilliland}, {Haas}, {Hall}, {Holman}, {Howard},
  {Howell}, {Isaacson}, {Jenkins}, {Klaus}, {Latham}, {Li}, {Marcy}, {Mazeh},
  {Quintana}, {Robertson}, {Shporer}, {Steffen}, {Windmiller}, {Koch}, \&
  {Borucki}}]{welsh2012}
{Welsh}, W.~F., {Orosz}, J.~A., {Carter}, J.~A., {et~al.} 2012, \nat, 481, 475

\bibitem[{{Wetherill}(1980)}]{weth1980}
{Wetherill}, G.~W. 1980, \araa, 18, 77

\bibitem[{{Wetherill} \& {Stewart}(1993)}]{weth1993}
{Wetherill}, G.~W., \& {Stewart}, G.~R. 1993, Icarus, 106, 190

\bibitem[{{Williams} \& {Cieza}(2011)}]{will2011}
{Williams}, J.~P., \& {Cieza}, L.~A. 2011, \araa, 49, 67

\bibitem[{{Windmark} {et~al.}(2012){Windmark}, {Birnstiel}, {Ormel}, \&
  {Dullemond}}]{windmark2012}
{Windmark}, F., {Birnstiel}, T., {Ormel}, C.~W., \& {Dullemond}, C.~P. 2012,
  \aap, 544, L16

\bibitem[{{Wisdom}(1980)}]{wisdom1980}
{Wisdom}, J. 1980, \aj, 85, 1122

\bibitem[{{Wisdom}(1982)}]{wisdom:1982}
---. 1982, \aj, 87, 577

\bibitem[{{Wisdom}(1983)}]{wisdom:1983}
---. 1983, Icarus, 56, 51

\bibitem[{{Xie}(2013)}]{xie2013}
{Xie}, J.~W. 2013, Acta Astronomica Sinica, 54, 79

\bibitem[{{Yorke} {et~al.}(1993){Yorke}, {Bodenheimer}, \&
  {Laughlin}}]{yorke1993}
{Yorke}, H.~W., {Bodenheimer}, P., \& {Laughlin}, G. 1993, \apj, 411, 274

\bibitem[{{Yoshida}(1990)}]{yoshida1990}
{Yoshida}, H. 1990, Physics Letters A, 150, 262

\bibitem[{{Youdin}(2011)}]{youdin2011b}
{Youdin}, A.~N. 2011, \apj, 742, 38

\bibitem[{{Youdin} \& {Chiang}(2004)}]{youdin2004a}
{Youdin}, A.~N., \& {Chiang}, E.~I. 2004, \apj, 601, 1109

\bibitem[{{Youdin} \& {Kenyon}(2013)}]{youdin2013}
{Youdin}, A.~N., \& {Kenyon}, S.~J. 2013, in Planets, Stars and Stellar
  Systems.~Volume 3: Solar and Stellar Planetary Systems, ed. T.~D. {Oswalt},
  L.~M. {French}, \& P.~{Kalas} (Dordrecht: Springer Science \& Business
  Media), 1

\bibitem[{{Youdin} {et~al.}(2012){Youdin}, {Kratter}, \& {Kenyon}}]{youdin2012}
{Youdin}, A.~N., {Kratter}, K.~M., \& {Kenyon}, S.~J. 2012, \apj, 755, 17

\bibitem[{{Zechmeister} {et~al.}(2013){Zechmeister}, {K{\"u}rster}, {Endl}, {Lo
  Curto}, {Hartman}, {Nilsson}, {Henning}, {Hatzes}, \& {Cochran}}]{zec2013}
{Zechmeister}, M., {K{\"u}rster}, M., {Endl}, M., {et~al.} 2013, \aap, 552, A78

\end{thebibliography}
\bibliographystyle{apj}

\begin{table}\footnotesize
\caption{\label{tab:kepler}
\kepler\ binaries. Along with orbital parameters
are estimates of the innermost stable orbit radius ($\acrit$)
and the forced eccentricity ($\eforce$).
}
\begin{tabular}{l|llcc|crc|cc}
\hline
\hline
\  & $\Mp\,(\Msolar)$ & $\Ms\ (\Msolar)$ & $\abin$\,(AU) & $\ebin$
& $a$\,(AU) & $e$\ \ \ \ & $r$\,(R$_J$) & $\acrit$\,(AU) & $\eforce$
\\
\hline
Kepler-16\tablenotemark{a} & 0.687 & 0.202 & 0.224 & 0.160 & 
 0.720 & 0.024  & 0.75 & 0.646 & 0.034
\\
Kepler-34\tablenotemark{a} & 1.049 & 1.022 & 0.228 & 0.521 & 
 1.086 & 0.209 & 0.76 & 0.833 & 0.002
\\
Kepler-35\tablenotemark{a} & 0.885 & 0.808 & 0.176 & 0.142 &
 0.605 & 0.048 & 0.73  & 0.496 & 0.002
\\
Kepler-38\tablenotemark{b} & 0.949 & 0.249 & 0.147 & 0.103 & 
 0.464 & $<$0.032 & 0.39 & 0.389 & 0.024
\\
Kepler-47\tablenotemark{c}  & 1.043 & 0.362 & 0.084 & 0.023 & 
 0.296 & $<$0.035 & 0.27 & 0.203 & 0.004
\\
PH1\tablenotemark{d}  & 1.528 & 0.378 &  0.174 & 0.212 & 
 0.634 & 0.054 & 0.55 & 0.527 & 0.044  
\\
Kepler-413\tablenotemark{e} & 0.820 & 0.542 & 0.099 & 0.037 & 
 0.355 & 0.118 & 0.39 & 0.253 & 0.003
\end{tabular}
\tablenotetext{a}{See \citet{doyle2011} and \citet{welsh2012};
orbital elements are from \citet[Table 1]{leung2013};}
\tablenotetext{b}{\citet{orosz2012b}.}
\tablenotetext{c}{\citet{orosz2012a}; A second planet ($r = 0.41$~R$_J$)
 is at $\sim 1$~AU.}
\tablenotetext{d}{\citet{schwamb2013} and \citet{kostov2013}; Kepler-64.}
\tablenotetext{e}{\citet{kostov2014}.}
\end{table}

\end{document}